\documentclass[review]{elsarticle}

\usepackage{amsmath,amssymb,amsfonts}
\usepackage{algorithmic}
\usepackage{graphicx}
\usepackage{textcomp}
\usepackage{xcolor}
\usepackage[ruled,linesnumbered]{algorithm2e} 
\usepackage[nolist]{acronym}
\usepackage{soul}
\usepackage{marginnote}
\usepackage{url}
\usepackage[inline]{enumitem}
\usepackage{tabularx}
\usepackage{amsthm}

\newcommand{\lastaccessed}{, last accessed Aug 12, 2022}

\newcommand{\myssec}[2]{%
  \subsection{#1}%
  \label{sec:#2}%
}
\newcommand{\rsec}[1]{%
  Sec.~\ref{sec:#1}%
}

\newcommand{\addeda}[1]{%
  {\color{blue}#1%
  }%
}
\newcommand{\addedb}[1]{%
  {\color{magenta}#1%
  }%
}
\newcommand{\addedc}[1]{%
  {\color{blue}#1%
  }%
}
\renewcommand{\addeda}[1]{#1}
\renewcommand{\addedb}[1]{#1}
\renewcommand{\addedc}[1]{#1}
\newcommand{\removed}[1]{%
  {\color{red}#1%
  }%
}
\renewcommand{\removed}[1]{}

\newcommand*{\thead}[1]{\multicolumn{1}{c}{\bfseries #1}}
\newcommand{\myfigeps}[3][width=3.1in]{%
\begin{figure}[tbp]%
\centering%
\includegraphics[#1]{figures/#2}%
\vspace{-1em}%
\caption{#3}%
\vspace{-1em}%
\label{fig:#2}%
\end{figure}%
}

\newcommand{\myfigfulleps}[3][width=\textwidth]{%
\begin{figure*}[tb]%
\centering%
\includegraphics[#1]{figures/#2}%
\caption{#3}%
\label{fig:#2}%
\end{figure*}%
}

\newcommand{\rfig}[1]{Fig.~\ref{fig:#1}}

\newcommand{\rtab}[1]{Table~\ref{tab:#1}}

\newenvironment{myinlinelist}%
{%
\begin{enumerate*}[label=(\roman*)]%
}%
{%
\end{enumerate*}%
}

{%
\begin{itemize}[parsep=0em,leftmargin=*,label={--}]%
}%
{%
\end{itemize}%
}

{%
\begin{enumerate}[parsep=0em,leftmargin=*,label=\arabic*.]%
}%
{%
\end{enumerate}%
}

\newcommand{\keypoint}[1]{\noindent\textit{\underline{Key point:} #1}}
\newcommand{\soadiff}[1]{\noindent\textit{\underline{Key difference:} #1}}

\begin{document}

\title{%
  FaaS Execution Models for Edge Applications
}

\author[1]{Claudio Cicconetti\corref{cor1}}%
\ead{c.cicconetti@iit.cnr.it}

\author[1]{Marco Conti}
\ead{m.conti@iit.cnr.it}

\author[1]{Andrea Passarella}
\ead{a.passarella@iit.cnr.it}

\cortext[cor1]{Corresponding author}
\address[1]{IIT, National Research Council, Pisa, Italy}

\begin{abstract}
  \addeda{In this paper, we address the problem of supporting stateful workflows
following a Function-as-a-Service (FaaS) model in edge networks.}
In particular we focus on the problem of data transfer, which can
be a performance bottleneck due to the limited speed of communication
links in some edge scenarios and we propose three different schemes:
a \textit{pure FaaS} implementation, \textit{StateProp}, i.e.,
propagation of the application state throughout the entire chain
of functions, and \textit{StateLocal}, i.e., a solution where the
state is kept local to the workers that run functions and retrieved
only as needed.
\addeda{%
We then extend the proposed schemes to the more general case
of applications modeled as Directed Acyclic Graphs (DAGs), which
cover a broad range of practical applications, e.g., in the
Internet of Things (IoT) area.
Our contribution is validated via a prototype implementation.
Experiments in emulated conditions show that applying the
data locality principle reduces significantly the volume of network
traffic required and improves the end-to-end delay
performance, especially with local caching on edge nodes and low link speeds.
}
\end{abstract}


\maketitle

  \section{Introduction}%
  \label{sec:introduction}%
  Computation offloading has been a trending topic in the networking
and cloud computing areas for some time now: it envisions that mobile
or resource-constrained devices offload part of their processing activities
to external entities with computation capabilities willing to undertake
the effort.
\addeda{A few years ago, the interest has then shifted towards the
edge of the network~\cite{Campbell2019}, as this enables latency-sensitive
applications that cannot afford a trip to far-away data centers.
However, recent cloud deployments already delocalize
data centers so that they are closer to the users~\cite{Mohan2020}:
research activities should not rely only on a closer-is-better-for-latency
motivation for edge computing, but rather look to the edge with a broader
perspective and find what it can realistically provide in specialized
use cases and applications.}

One such opportunity that is emerging is to employ at the edge a \ac{FaaS}
model~\cite{Aslanpour2021}: the application is decomposed into functions
that are invoked individually or in a chain.
\ac{FaaS} is very well suited to many \ac{IoT} applications of
practical interest for what concerns the programming model (functional
event-based), an efficient utilization of resources (both at device-
and edge node-level) and the promises of high scalability.
The latter stems from symbiosis with a \textit{serverless computing}
framework, where functions are invoked in containers that are
orchestrated in a highly flexible virtualization
infrastructure~\cite{Khandelwal2020}.
\addeda{Results have demonstrated that serverless is more suitable than
a microservice architecture for unpredictable requests accompanied by
a large size of the response, due to the scaling agility~\cite{Fan2020}.}
Serverless/\ac{FaaS} are major trends in cloud
computing~\cite{Castro:2019:RSC:3372896.3368454}, thus edge
deployments/applications could benefit from the ample availability
of industry-grade commercial and open source solutions~\cite{Yussupov2020},
even though some advances beyond the state of the art are required to
make a good use of resources in this different environment~\cite{Xie2021}.

In particular, serverless relies on the implicit assumption that
the container location is irrelevant to performance, which has led
to a \textit{stateless} \ac{FaaS} paradigm: applications that have a
state rely on external services offered (and billed separately) by
the cloud provider~\cite{Hellerstein2018}.
\addeda{In \cite{Eismann2021a} the authors have analyzed open-source
and proprietary datasets and they have found that only 12\% of the
serverless applications in production are \textit{truly} stateless,
whereas the others rely on managed services such as storage (61\%)
and databases (48\%).}
In the cloud, this may lead to a slightly sub-optimal utilization
of the resources, e.g., due to the inability to keep hot content
in caches~\cite{Shahrad2019}, but at the edge the impact becomes
much more ominous: here the cost of transferring data between
function executors and external services is, in general, much higher
than in a data center, hence the network may easily become a limiting
factor~\cite{Rausch2021}.
Furthermore, one of the most appealing features of \ac{FaaS} is the
opportunity for the service provider to
\addedb{compose applications as complex workflows of invocations,
so that the output of a function is not returned immediately to the
user but delivered to one (or more) successors for further processing.
}%
This exacerbates the above implications of location dependency,
\addedb{which become crucial when migrating from a microservice to
a serverless architecture~\cite{Jin2021}.
}

In this paper, we address the problem of data transfer
(including both arguments/return values and the application state)
within a workflow of stateful functions,
\addedb{
motivated by a practical use case illustrated in \rsec{motivation}.
Then, after reviewing the state of the art (\rsec{soa}), we summarize
in \rsec{contribution} the findings in our previous work \cite{Cicconetti2021}, where we have
proposed three fully decentralized execution models for applications that
can be modeled as chains of functions\addedc{: PureFaaS, StateProp, and StateLocal}.
The execution models are extended to the more general
case of applications modeled as \acp{DAG} in \rsec{dag}.
}
\addedc{As we will see later, PureFaaS, which is closest to state-of-the-art
serverless platforms, is surpassed by StateProp and StateLocal, which
achieve reduced traffic and smaller delays.
However, they incur the cost of a slightly higher system complexity,
because they require the ability to embed the application's state
as function arguments (StateProp) or keep the state at edge nodes
(StateLocal), in addition to a more profound knowledge of the
application workflow (\ac{DAG} of function invocations and state
dependencies).}
\addeda{
We have implemented a prototype of the proposed solutions, which
we use to compare their performance in emulated network experiments
(\rsec{prototype}).
%
%
We draw the conclusions in \rsec{conclusions}.
%
}%

\addedb{%
  \section{Motivation}%
  \label{sec:motivation}%
  In this section we describe a motivating example, inspired from the
activity ongoing in the H2020 MARVEL collaborative R\&D
project\footnote{\url{https://www.marvel-project.eu/}\lastaccessed},
co-funded by the European Commission, in which we participate.
The project defines a framework for real-time analytics in smart
cities, addressing several applications of high impact to
citizens validated in two pilots, municipality of Trento
and public streets in Malta: automated detection of
anomalous traffic conditions, monitoring of crowded areas, 
protection of vulnerable users (pedestrians, cyclists) in street
junctions, emotion recognition in public events, and many others.

\myfigfulleps{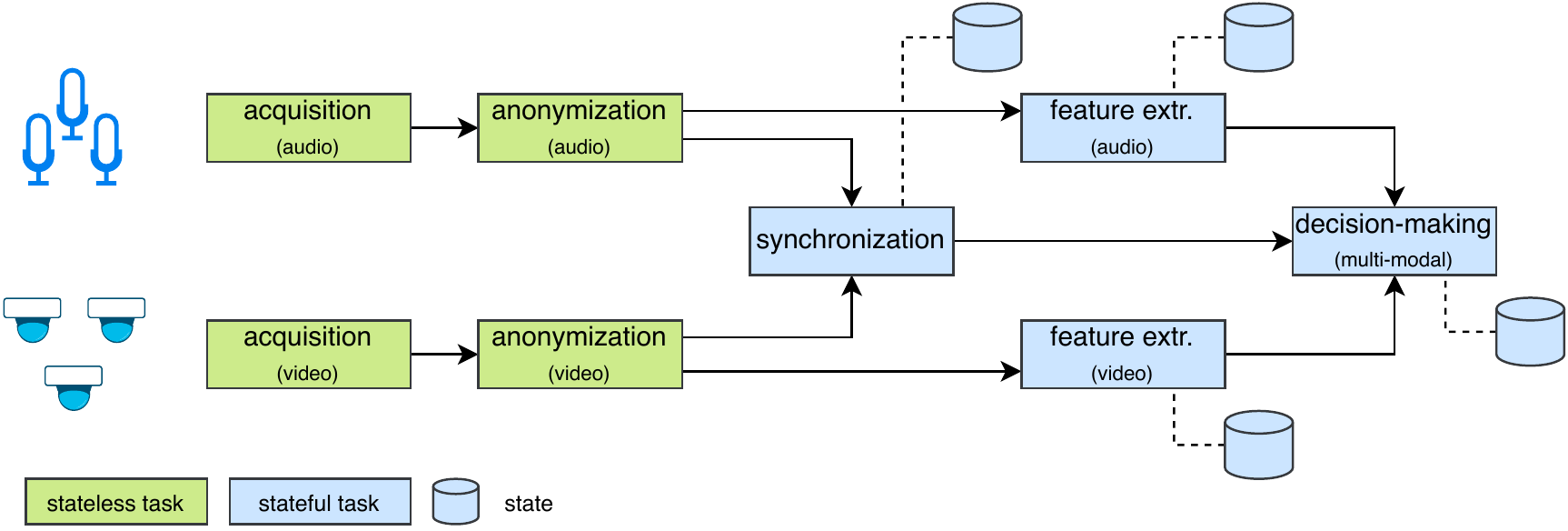}{%
Archetypal smart city application for audio-video analytics in the MARVEL project.}

All the applications have the same general structure illustrated in
\rfig{problem-marvel}: starting from the acquisition of data from sensors,
first there is an anonymization phase to remove
personal data, then the relevant features are extracted and used to trigger
a \ac{ML}-driven decision-making process.
\addedc{Despite its potential benefits, \textit{vanilla serverless computing
cannot be adopted here} for two reasons.
First, a stateless execution is not sufficient: (some of) the components require
read/write access to a
per-application state, e.g., with video streams to cross-correlate
the current frame with previous ones in a window.
Second, \ac{FaaS} platforms support chain of functions but our applications
have multiple sources per application (cameras + microphones),
which requires the workflow to be designed
as a \ac{DAG} for synchronization and fusion purposes.}

\myfigeps{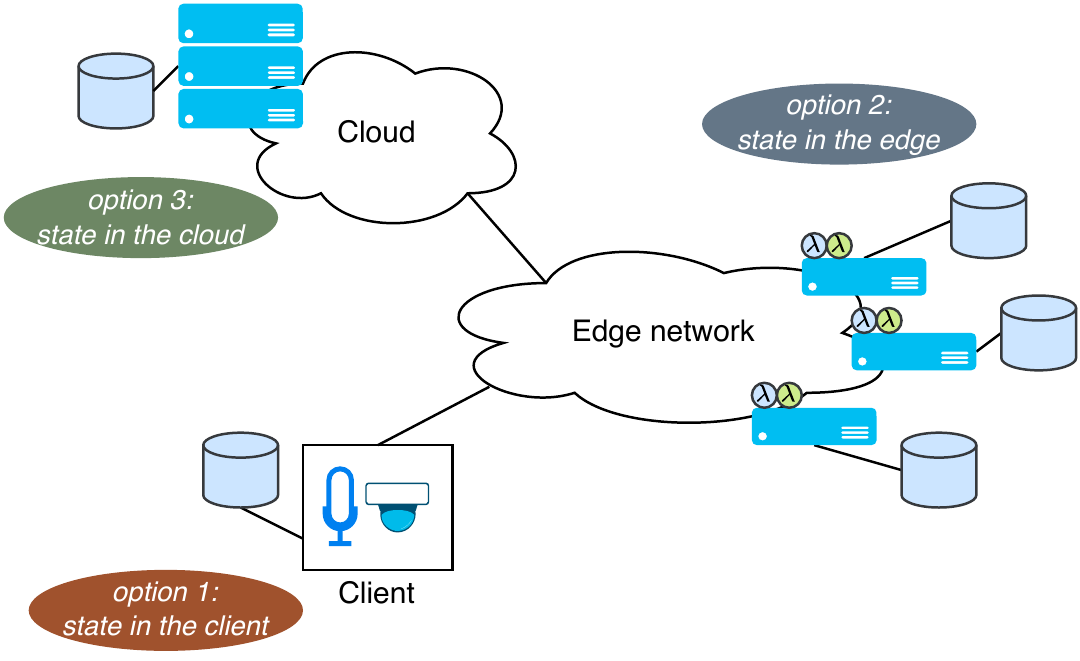}{%
Architecture options on where to keep the application's state:
\ul{option~1}: the state remains in the client, which embeds it in the function
arguments and return value so that the execution remain stateless;
\ul{option~2}: the state is kept by the edge nodes; 
\ul{option~3}: the state is handled by a cloud service, which is the default today.}

\addedc{In this paper we address both these aspects:
in \rsec{contribution} we discuss the possible options on where to maintain the
application's state, summarized in \rfig{problem-marvel-sim}, while
the support of \acp{DAG} is illustrated in \rsec{dag}.
In \rsec{prototype} we evaluate our proposed system with a prototype
implementation in a mininet testbed, which also used for the motivating
example in this section.}.

%

\myfigeps{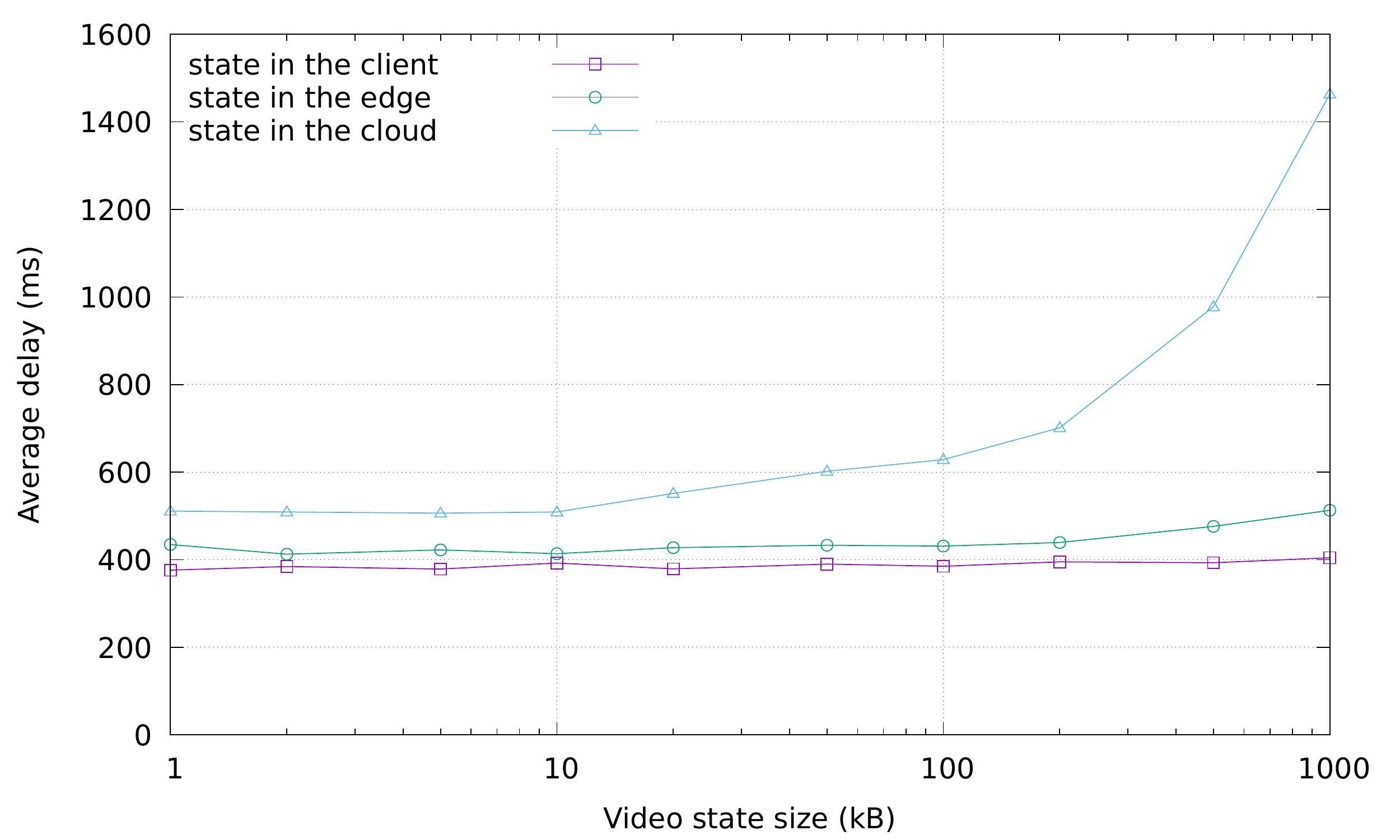}{%
Average delay of the workflow in \rfig{problem-marvel} with the
three options in \rfig{problem-marvel-sim}.
The edge links have 1~Gb/s bandwidth with 1~ms
latency, whereas the cloud node is connected through a 10~Mb/s link
with 20~ms latency. 
The full details on the methodology and tools used are in \rsec{prototype}.}

In \rfig{402-avg-delay} we show the average delay of the workflow when
increasing the state size of the feature extraction video function
from 1~kB to 1~MB; the state of the corresponding function for audio is 1/10
of the latter, those of the other functions 1/100.
The state sizes used here are arbitrary but representative of real
use cases.
As can be seen, when the state is kept in the cloud the average
delay is always higher than that in the other cases, and it grows
significantly as the state size increases.
On the other hand, there is no noticeable increase when the state is maintained
at the edge or in the client, with the latter exhibiting the smallest delay
(though at the cost of higher network traffic, not shown).

The results found in this motivating example show that the performance
of data-intensive applications made of \ac{DAG} stateful functions,
such as real-time smart city analytics, significantly depends on
where the state is kept, which is the subject of this work.
It is worth mentioning that these applications are not a special case.
For instance, also robotic applications are typically designed using
a \ac{DAG} model~\cite{Alirezazadeh2021} and, in general, we have
analyzed the traces of real-life cloud applications collected from
a production system in an Alibaba data center\footnote{\url{https://github.com/alibaba/clusterdata/blob/master/cluster-trace-v2018/trace_2018.md}\lastaccessed.}
and we have found that a non-negligible fraction of applications
consist of \acp{DAG}: 21.7\%, with single tasks being
28.6\% and chains 49.7\%.

}

  \section{Related work}%
  \label{sec:soa}%
  \addeda{
Serverless platforms in the cloud hinge on the underlying container
orchestration systems, which handle autoscaling and are responsible for
consistent performance.
However, these orchestration tools are inefficient when used at the edge,
where devices are heterogeneous and clustered, which
causes sub-optimal performance~\cite{Carpio2020}.
%
%
For a comprehensive review on all the aspects of resource management in
serverless systems we refer the interested reader to the survey
in~\cite{Mampage2021}.
In the following, we focus on some works that are especially relevant
to our contribution\addedb{, that is the definition of suitable
execution models to handle chain/DAG composition of stateful functions
in serverless edge networks.}
}


At the edge, the problem of data locality has
been introduced neatly in~\cite{Rausch2021}, where the authors
%
have proposed to influence resource scheduling in \ac{K8s}
by adjusting its internal weights based on metadata specified by
the application developers.
%
The more abstract problem of directly allocating
functions to edge nodes for complex applications consisting of
multiple inter-related functions has been studied in~\cite{Wang2020b}
with a mathematical formulation, which takes into
account different categories of cost (i.e., activation, placement,
proximity, sharing).
%
\addeda{The problem has been also studied in the context of Mobile
Edge Computing (MEC) with \ac{DAG}-modeled applications in
\cite{Liao2021}, where the authors sought the goal of meeting as many
application request deadlines as possible using an online algorithm
approximating an NP-hard scheduling problem.
Finally, in~\cite{Nicolaescu2021} the authors have proposed \textsc{SEND}, an
in-network storage management system realized through edge-deployed data
repositories, which places intelligently raw and processed data
based on locality or popularity criteria.
}

\soadiff{
    Our work is complementary to the above studies because we
    do not address the allocation of containers/tasks to edge nodes,
    but rather propose solutions to manage the applications' state in
    \ac{FaaS} under a given allocation.
}



Supporting stateful applications is one of the key research
challenges identified in the position paper~\cite{Khandelwal2020} for
serverless computing in the cloud.
%
%
\addeda{A datastore for edge computing with consistent replicas has
been also proposed in~\cite{Mortazavi2020}, which reconciles only
the data that are relevant to a given session for performance
reasons.
In~\cite{Wu2020} the authors have proposed \textsc{HydroCache}, a distributed
data caching system with multi-site causal consistency, which can be
used as state management for serverless \ac{DAG} functions and has been
shown to outperform uncached platforms in the cloud.
Fault tolerant function execution, with embedded garbage collection, has been
addressed by the authors in~\cite{Zhang2020} and found to be both effective
and affordable on AWS Lambda.
Finally, Boki~\cite{Jia2021a} has been proposed as a serverless platform
that enables stateful applications via shared logs, which have
ordering, consistency and fault tolerance properties.

\soadiff{
    Currently stateful \ac{FaaS} \textit{in edge
    networks} is largely unexplored, which is a motivation for our work
    to explore different approaches for argument and state
    distribution with chains and \acp{DAG} of function invocations.
    In a practical deployment, sophisticated state management systems
    can be used in combination with the execution models that we propose
    in this work.
    }
}

\addeda{The need for efficient serverless platforms that can scale down to small
edge devices is illustrated in~\cite{Hetzel2021}, which proposes to integrate
a computation model based on the actor pattern with content-based networking.
To address also microcontrollers, the authors have used a pub-sub messaging
system underneath.
A similar approach has been followed in \cite{Shillaker2020}, which presents
\textsc{Faasm}, where user-space isolation abstraction is provided via the
use of the WebAssembly run-time environment and applications can share
state using a hierarchical \ac{KVS}.

\soadiff{
    The actor model has interesting similarities with serverless.
    As a matter of fact, the integration of WebAssembly platforms using
    the actor pattern with \ac{FaaS} platforms is under way (e.g.,
    wasmcloud\footnote{\url{https://wasmcloud.dev/}\lastaccessed} and
    OpenFaaS\footnote{\url{https://www.openfaas.com/}\lastaccessed}).
    We believe our work makes a valid contribution across both
    domains, as we propose different execution models for stateful
    execution of chain/\ac{DAG} workflows, which can be combined or
    tailored to the specific needs of the scenario and applications.
    }
}

\addeda{Some recent works have focused on the specific issue of
resource management for \ac{DAG} workflows in serverless platforms.
In~\cite{Carver2019} the authors have identified a set of techniques
to make \ac{DAG} schedulers aware of the serverless platform they are
running on, tested on AWS Lambda.
The opposite approach has been followed in~\cite{Bhasi2021}, where
the authors have defined an orchestration framework to
match the application performance requirements via appropriate
provisioning of containers in a K8s cluster.

\soadiff{
    These works contribute to the motivation of our research activity,
    as they deal with applications that can be modeled with \acp{DAG},
    which however remains complementary: we focus on different schemes to
    pass on the arguments and states of the application, which is a
    different (though possibly related) problem to container resource
    management.}
}

\addeda{Finally, we mention that practical applications might also
require mechanisms for the realization of patterns beyond the execution
of stateful \acp{DAG}.
Microsoft's commercial serverless platform, \acp{ADF}, also allows
to define critical sections for the atomic execution of some functions
in a workflow~\cite{Burckhardt2021}, whereas explicit parallelization
of function execution has been investigated in~\cite{Zhang2020a}.

\soadiff{
    We recognize that some applications have specific needs that cannot
    be addressed efficiently by a single solution.
    In this work we have focused on chain and \ac{DAG} workflows, which
    are very common and already cover a broad range of applications of
    practical interest, and we leave for future work further specializations,
    including critical sections and explicit parallelism.
    }
}%

  \section{Stateful function chains}%
  \label{sec:contribution}%
  \addedb{%
In this section we introduce the system model and notation used in
the paper and we summarize the findings in \cite{Cicconetti2021},
which tackled the execution of chains of functions on serverless
platforms deployed at the edge, which are extended to the more
general case of \ac{DAG}-modeled applications in the next section.
We conclude the section with considerations about confidentiality
in \rsec{cont:ipr}.
}

%
%
%


\myfigeps[width=\textwidth]{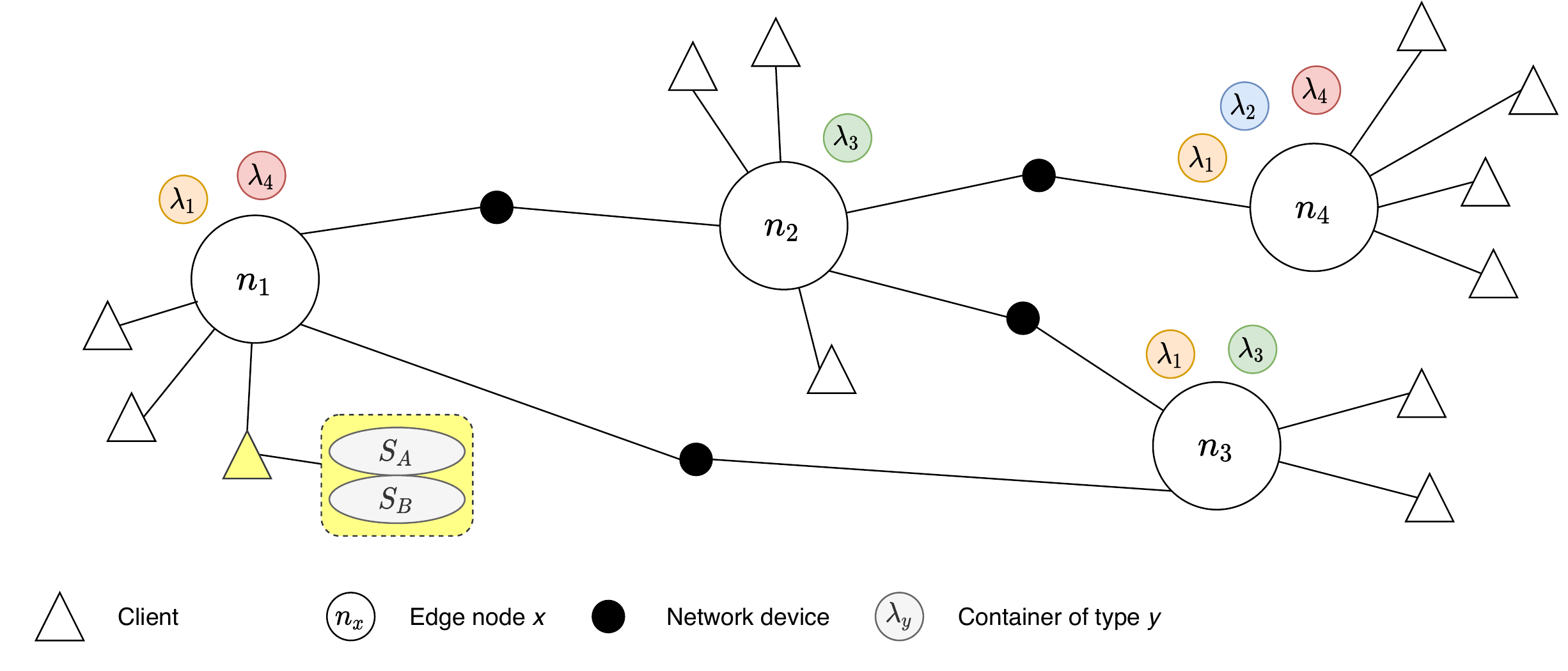}{System model.}

The system model is illustrated with the help of the example in
\rfig{problem-model}.
%
In the figure we have four edge nodes indicated as $n_i$, each
hosting a serverless platform that can execute lambda functions of one or
more types, indicated as $\lambda_i$, via a pool of \textit{workers}.
For instance, $n_1$ can execute lambda functions $\lambda_1$ and $\lambda_4$
but not $\lambda_2$ and $\lambda_3$.

\myfigeps{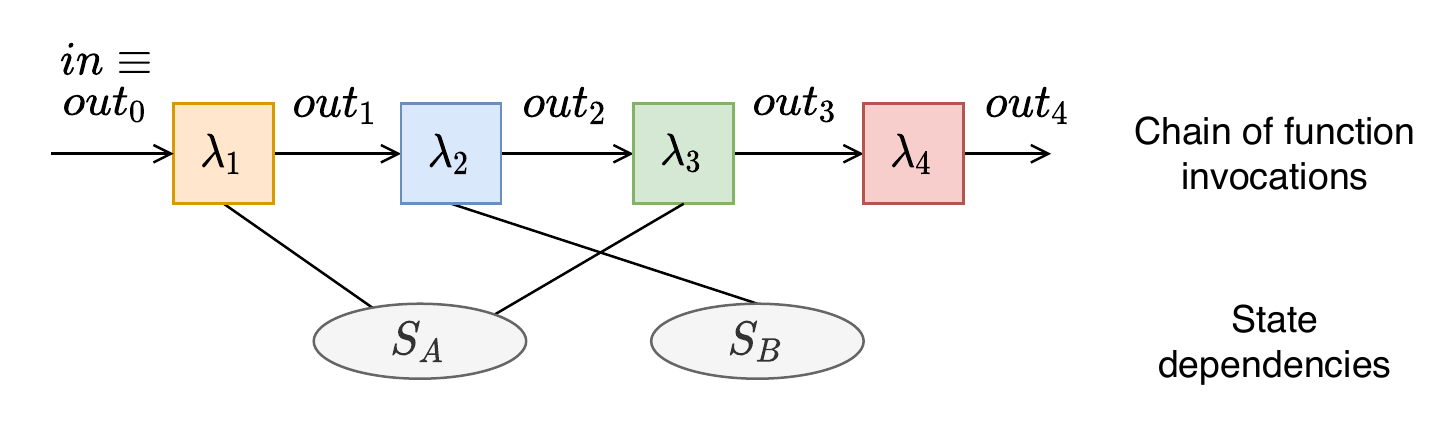}{Example chain application.}

Let us now consider the example user application depicted in
\rfig{problem-chain}: the client needs the input to be provided to
$\lambda_1$, which also requires (and possibly modifies) the
application state $S_A$, whose output $out_1$ needs to be provided
to $\lambda_2$, also requiring access to $S_B$, which feeds $\lambda_3$
and so on, until the final output $out_4$ is returned to the client.
The target application is colored in yellow and represented running
on the client device, also showing its two states $S_A$ and $S_B$.
\addedb{
In the cloud, \textit{stateful} functions are realized by means of
\textit{stateless} functions that access external services, such
as in-memory databases or storage services.
However, this approach is not efficient at the edge, as shown in
\rsec{motivation}.
Our alternatives are to keep the state in the client vs.\ in the edge nodes.
In \cite{Cicconetti2021} we have explored three different approaches, which
are summarized below: PureFaaS and StateProp (the state remains in the client)
and StateLocal (the state is kept by the edge nodes).
In the following we assume that the allocation of functions to nodes is:
$\left[\lambda_1,\lambda_2,\lambda_3,\lambda_4\right]
\rightarrow \left[n_1,n_4,n_2,n_1\right]$.
}



\myfigeps{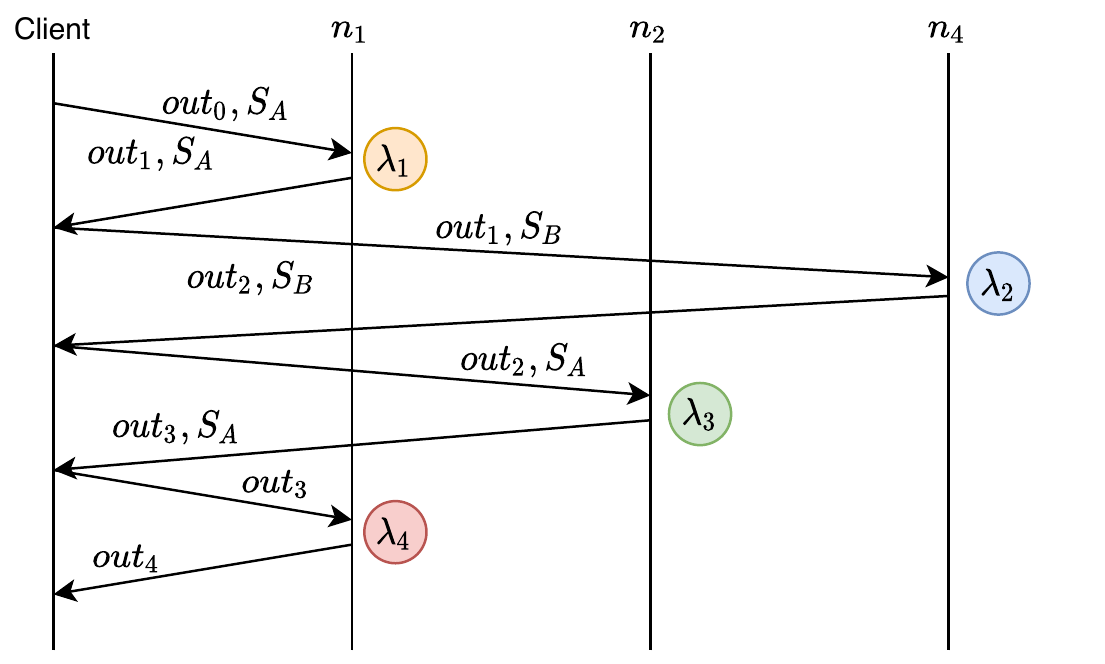}{%
PureFaaS execution diagram of the application in \rfig{problem-chain}
at the edge in \rfig{problem-model}.}

With \textbf{PureFaaS}, the functions in the chain are executed one
after another, and the required state of each function is transferred
back and forth with every invocation, as illustrated in
\rfig{flow-diagram-1a}.
%
%
%
This strategy can be easily realized on commercial/open source
serverless platforms provided that:
\begin{myinlinelist}
    \item the signature of the function (both arguments and return
    value) supports the client embedding the required
    state\footnote{\addeda{Commercial platforms may limit the amount
    of data that can be embedded into function
    invocations~\cite{GarciaLopez2019}: for instance, with AWS this
    limit is a mere 32~KB, whereas with IBM it is 5~MB, but the
    overhead has been shown to increase non-linearly with the
    arguments' size.
    Only with Microsoft's \acp{ADF} it seems there is no theoretical
    limit, but a compression mechanism is triggered automatically above
    60~KB.}};
    \item the client is aware \textit{a priori} of the state that will be needed
    by every next function invoked.
\end{myinlinelist}
%
%



\myfigeps{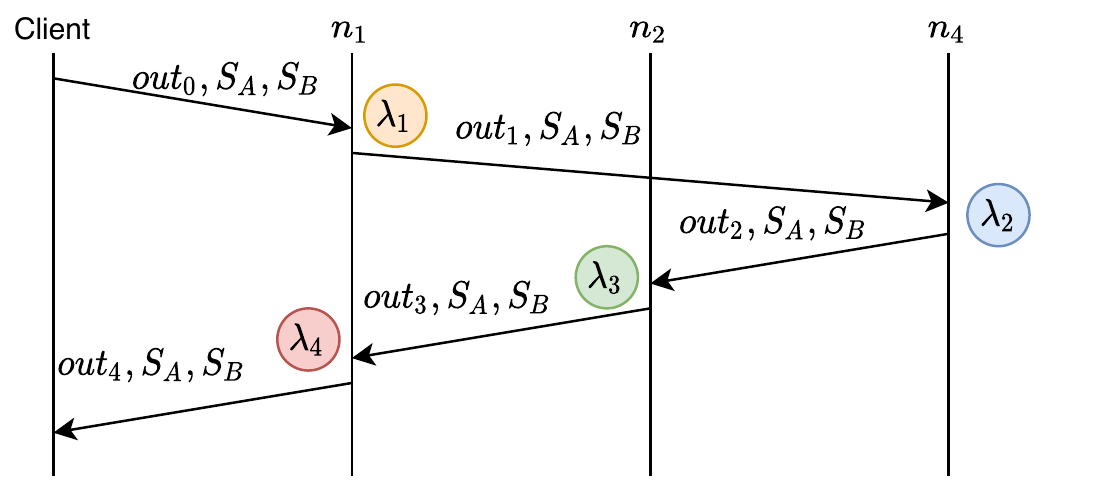}{
StateProp execution diagram of the application in \rfig{problem-chain}
at the edge in \rfig{problem-model}.}

\addedb{
\textbf{StateProp} is similar but it makes use of the chaining capability
made available by most serverless platforms.
As shown in \rfig{flow-diagram-1b}, the client embeds the full state of the
application into the function arguments and return values:
a function that does not use the embedded state will simply let it pass through,
while the others will embed as function arguments the modified state received,
which will eventually be returned to the client.
}

\myfigeps{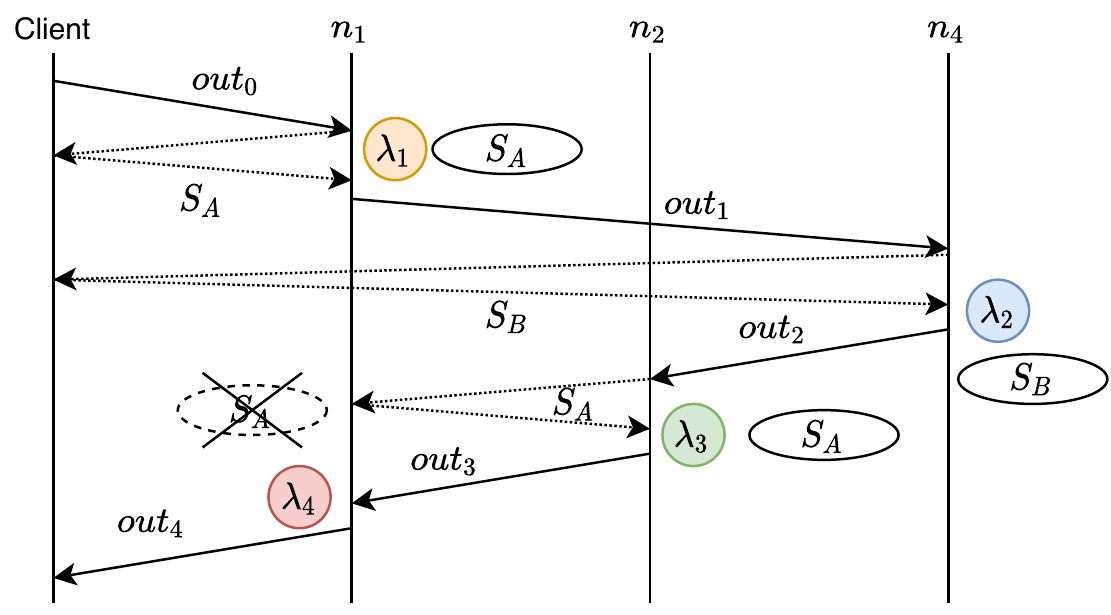}{
StateLocal execution diagram of the application in \rfig{problem-chain} at the edge in \rfig{problem-model}.}

\addedb{
Finally, \textbf{StateLocal} keeps the state in the edge nodes
as illustrated in \rfig{flow-diagram-1c}: rather than embedding the state
in the function invocations, only \textit{pointers} are passed.
When a lambda function needs a state, it retrieves it via the pointer, and
then it becomes its new owner, thus modifying the state's pointer in the
subsequent function invocation along the chain.
This way, the client will be eventually returned the list of updated pointers
to all its states, to use them in subsequent application executions or to
withdraw the states from the edge nodes, if ever needed.
}

\myssec{Disclosure of proprietary information}{cont:ipr}

Function composition in any serverless platform introduces the risk of
disclosing proprietary information about the application's logic to the
platform provider: even though the implementation of a single function can
remain private, as only the end-points are needed for the sake of function
invocation, some information about the algorithms being executed
could be deduced by the way the functions are chained and their usage
patterns.
This risk becomes even greater with \ac{DAG} applications made of elementary
building blocks, as their richer expressivity could offer further insights
about the overall service logic.
While we recognize that there can be use cases where disclosing this
minimal amount of information to a third party (and potentially a competitor)
may not be deemed acceptable, we believe such a risk cannot be considered
a show-stopper in the majority of practical scenarios.
Therefore, we defer the investigation of the issue to future works in this area.%


\addeda{%
  \section{Extension to DAGs}%
  \label{sec:dag}%
  In this section we extend the execution models in \rsec{contribution}
to applications that can be modeled as \acp{DAG}.
%
%
We introduce \ac{DAG}-specific notation in \rsec{dag:consistency},
then address the extension of the stateful execution models in the
previous section separately for PureFaaS (\rsec{dag:purefaas}) and
StateProp/StateLocal (\rsec{dag:prop}).


\myssec{State consistency}{dag:consistency}

An application modeled as a \ac{DAG} consists of a set of tasks
\addedb{(we use the terms \textit{tasks} and \textit{functions}
interchangeably)} with precedences: an edge $\lambda_i \rightarrow
\lambda_j$ exists if task $\lambda_i$ must be executed before task
$\lambda_j$.
\addedb{%
The set of precedences define a directed graph, called \textbf{task dependency
graph}, which cannot contain cycles by definition (recall the `A' in \ac{DAG}
stands for Acyclic), otherwise the execution would never end.
%
%
%
Dependencies can be of the input/output type, i.e., $\lambda_i
\rightarrow \lambda_j$ means that the output of task $\lambda_i$
is needed by task $\lambda_j$, or a means of synchronization like
in a message-passing system, the distinction is irrelevant to our
purposes.
}%
To keep the notation consistent with \rsec{contribution}, we
indicate with $out_i$ the output of task $i$, even though we note that in
a \ac{DAG} there could be \textit{multiple}, possibly different, outputs
for each task.
Support of different task outputs is a mere implementation detail,
which does not affect the contribution illustrated in this section, except
for a trivial generalization of the derivations.

\myfigeps[scale=0.5]{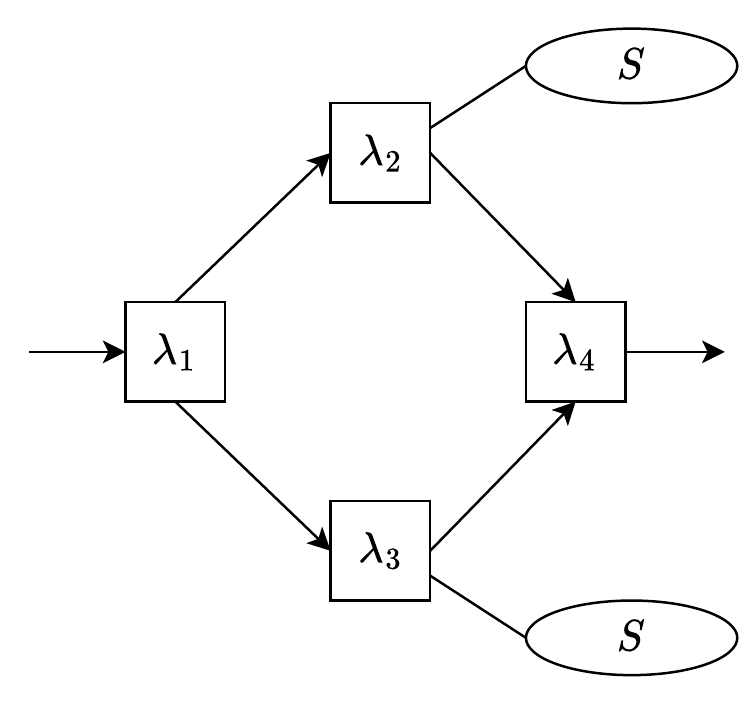}{Task-state dependency graph of an example application.}

\addedb{%
In our work we address workflows make of \textit{stateful functions},
thus any task may also depend on some states.
As in \cite{Cicconetti2021} (summarized in \rsec{contribution}), we capture
the stateful nature of functions via the \textbf{state dependency graph},
which is an undirected graph where edge $\lambda_i \rightarrow S_x$ means
that the task $\lambda_j$ needs to access state $S_x$.
The union of the task dependency graph and the state dependency graph
produces the \textbf{task-state dependency graph}; an example is
illustrated in \rfig{dag-1} for an application made of four tasks, two
of which ($\lambda_2$ and $\lambda_3$) use state $S$.
}

\myfigeps{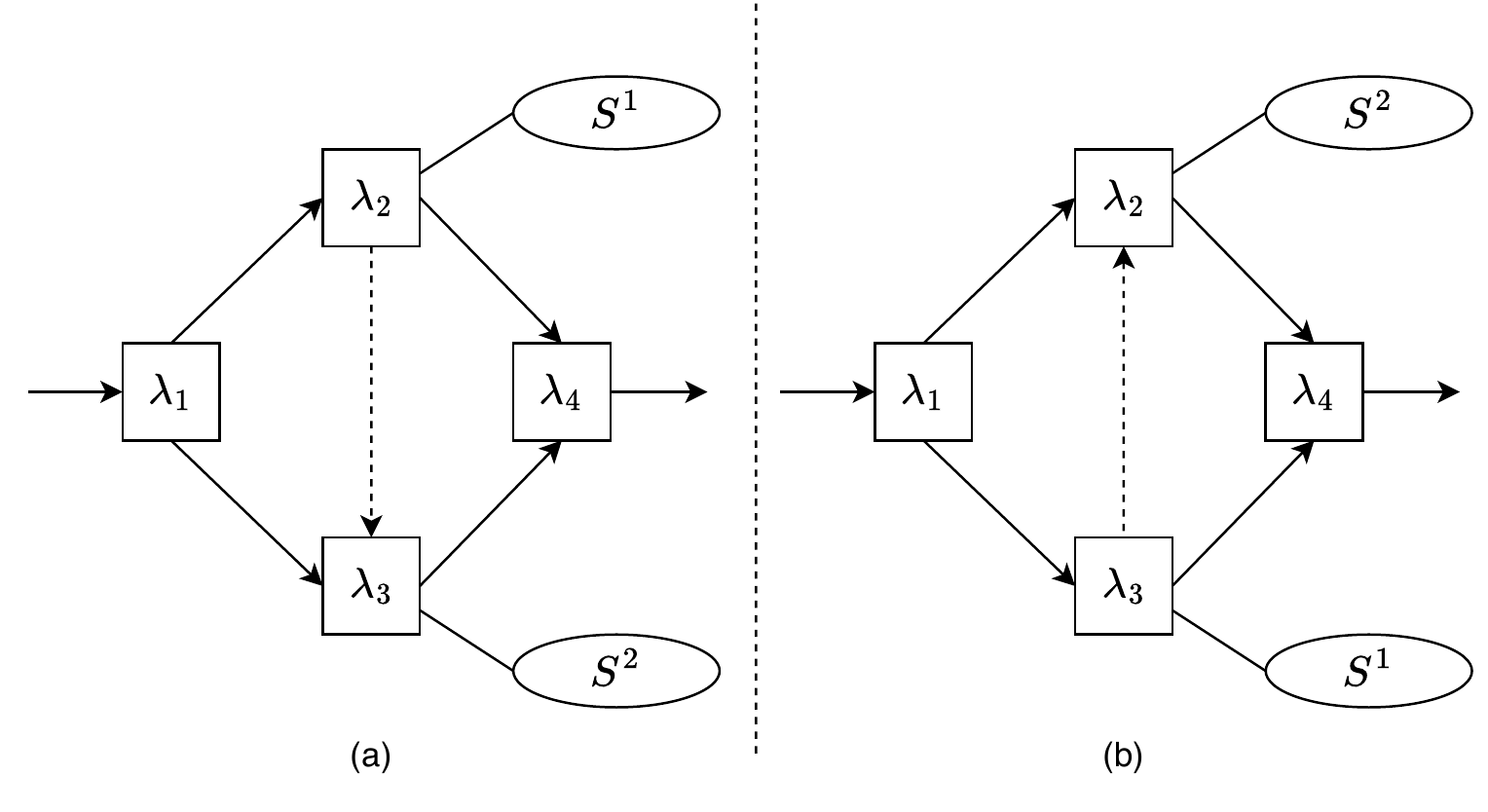}{Virtual links added between tasks $\lambda_2$ and $\lambda_3$,
both depending on state $S$.}

\addedb{%
It can happen, like in the example in \rfig{dag-1}, that multiple
tasks need to operate on the same state during a single execution
of the application.
Depending on the internal logic of the application, it can happen that:
}%
%
%
\begin{myinlinelist}
    \item the order of execution does not matter;
    \item $\lambda_2$ must be executed before $\lambda_3$;
    \item $\lambda_3$ must be executed before $\lambda_2$.
\end{myinlinelist}
\addedb{%
The serverless platform needs to know from the application the
temporal order of execution of tasks that depend on a shared state
to maintain the \textbf{causal consistency of the states}.
We then capture such temporal order by augmenting the task-state
dependency graph as follows: a virtual edge $\lambda_i \rightarrow
\lambda_j$, represented as a dashed line in the figures below, is
added if $\lambda_i$ and $\lambda_j$ both use the same state and
$\lambda_i$ must be executed before $\lambda_j$ to guarantee causal
consistency of the shared state.
}%
In the previous example, this means adding an edge
$\lambda_2\,\rightarrow~\lambda_3$ if $\lambda_2$ has to be executed
first (case `a' in the figure) vs.\ $\lambda_3~\rightarrow~\lambda_2$
if $\lambda_3$ has to be executed first (case `b' in the figure).
\addedb{In this section we assign a superscript to states shared by multiple
tasks to express the temporal dependency order: in \rfig{dag-2} $S^1$ must be accessed
first, $S^2$ second.}
%

\myfigeps{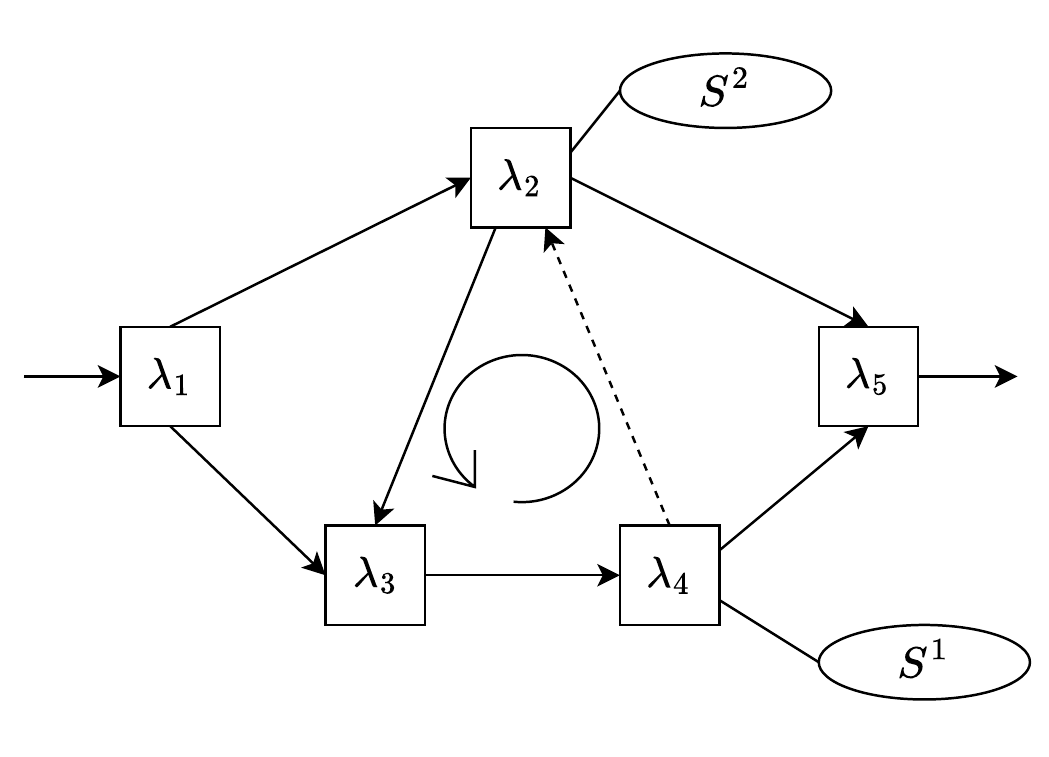}{%
Causal consistency induces a cycle in the augmented task-state dependency graph.}

The addition of the virtual edges affects the \textbf{parallelism}
that can be achieved: without state dependencies the
\ac{DAG} in \rfig{dag-1} can executed as $\{ \lambda_1, \lambda_2|\lambda_3,
\lambda_4$ (where $|$ means that the two tasks can be executed in parallel),
but with that in \rfig{dag-2} this is not possible.
\addedb{%
Furthermore, in order for the augmented task-state dependency graph to remain
well-formed, no cycles must exist, also including the virtual edges,
i.e., the state-induced temporal order dependencies.
}
%
Let us consider for instance the application in \rfig{dag-3}.
\addedb{The virtual edge $\lambda_4 \rightarrow \lambda_2$ would
be needed, however this would create the cycle $\lambda_2 \rightarrow
\lambda_3 \rightarrow \lambda_4$, which in turn leads to a
\textit{deadlock}:
$\lambda_4$ cannot be executed until it receives the
output of $\lambda_3$, which in turn cannot run until it receives the
output of $\lambda_2$, which cannot access state $S$ before the execution 
of $\lambda_4$ is complete.}
Since this kind of situations can be detected by the application, and
they reflect a logic design issue, we assume hereafter that our
applications of interest are only those with an acyclic augmented
task-state dependency graph.

\keypoint{With stateful \ac{DAG} applications, the causal consistency of
the execution must be guaranteed.
We propose to do so by defining, for each state, the order in which the tasks
depending on it will be executed, as reflected by virtual edges
in the augmented task-state dependency graph.
}

\myssec{Pure Faas}{dag:purefaas}

\myfigeps{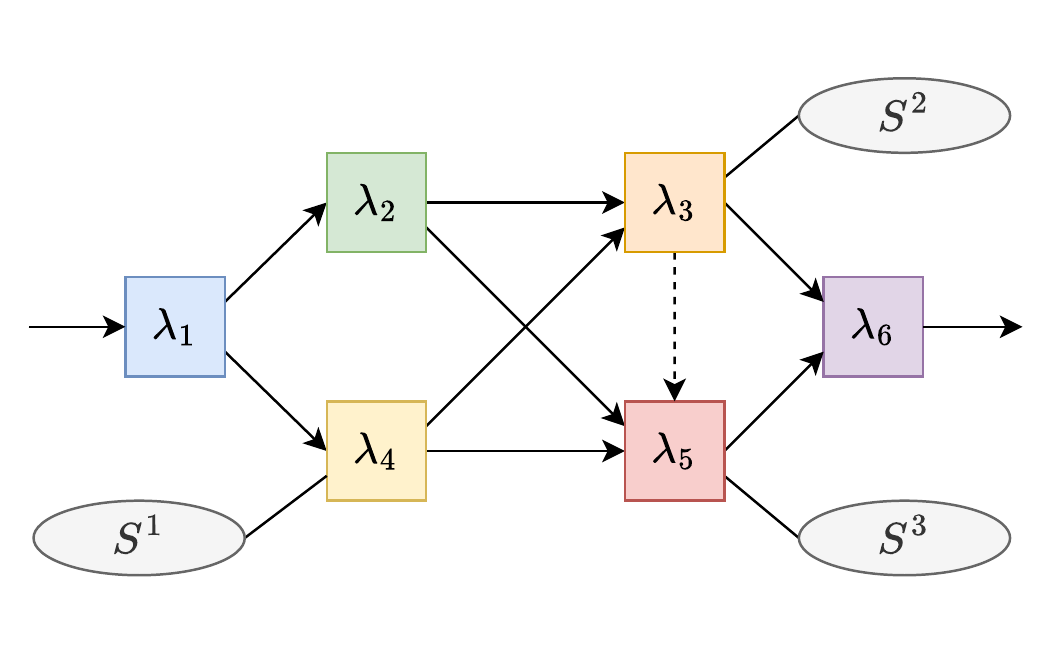}{Augmented task-state dependency graph of an example application
used to illustrate the extension of the execution models proposed in
\rsec{contribution} to the case of \ac{DAG}.}

The extension of PureFaaS approach is straightforward.
At each step, there is a set of callable functions, which are all those
whose task-state dependencies are verified.
The client can call them in parallel or in sequence (order is arbitrary)
depending on its internal logic and capabilities.

\myfigfulleps{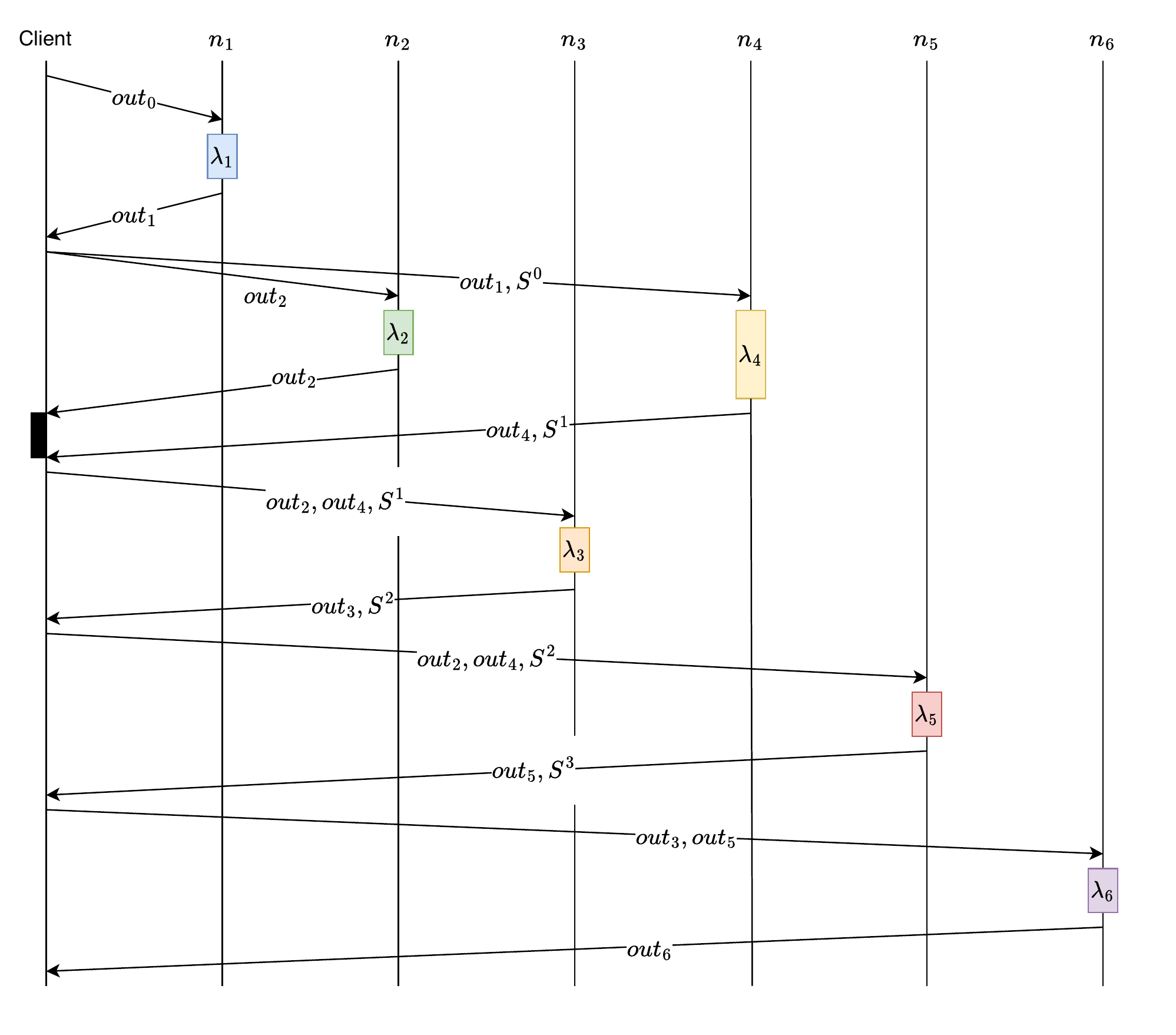}{Execution of
the example \ac{DAG} application in \rfig{dag-4} with PureFaaS.}

To better illustrate our point, we make use of the example application
in \rfig{dag-4}.
The sequence diagram with PureFaaS is shown in \rfig{dag-5}, where
we assume for better readability that function $\lambda_i$ is always
executed on edge node $i$.
\addedb{Moreover, with a slight overload of notation, with $S^i$
we not only indicate the temporal order dependency of the state $S$
in the augmented task-state dependency graph (as defined in
\rsec{dag:consistency}), but we also refer to its subsequent
modifications: $S^0$ is the initial state before the workflow
invocation, $S^1$ its modified version after the execution of
$\lambda_4$ (which connects to $S^1$ in the graph), and so on until
the final version $S^3$ of the state at the end of the workflow.
}%
As can be seen, the tasks $\lambda_2$ and $\lambda_4$ are executed in parallel,
but the client has to wait for the slowest of the two (in the example:
$\lambda_4$) before it can continue: this waiting time is represented in
the diagram with a black rectangle.
Apart from the opportunity to parallelize some tasks, there is no other
change with respect to PureFaaS when used with chains of functions invocations.
%

\keypoint{PureFaaS remains the same with chains and \acp{DAG}.}

\myssec{StateProp/StateLocal}{dag:prop}

On the other hand, StateProp and StateLocal cannot be used with \ac{DAG}
applications without modifications like PureFaaS.
There are different reasons for this, which we will explain hereafter.
Briefly, we recall that StateProp/StateLocal both rely on the worker invoking
the next function as the current one is complete; they differ on the way they
manage the state: StateProp carries it along the function invocation chain,
whereas StateLocal keeps it within the edge node that last used it.

\myfigeps[scale=0.5]{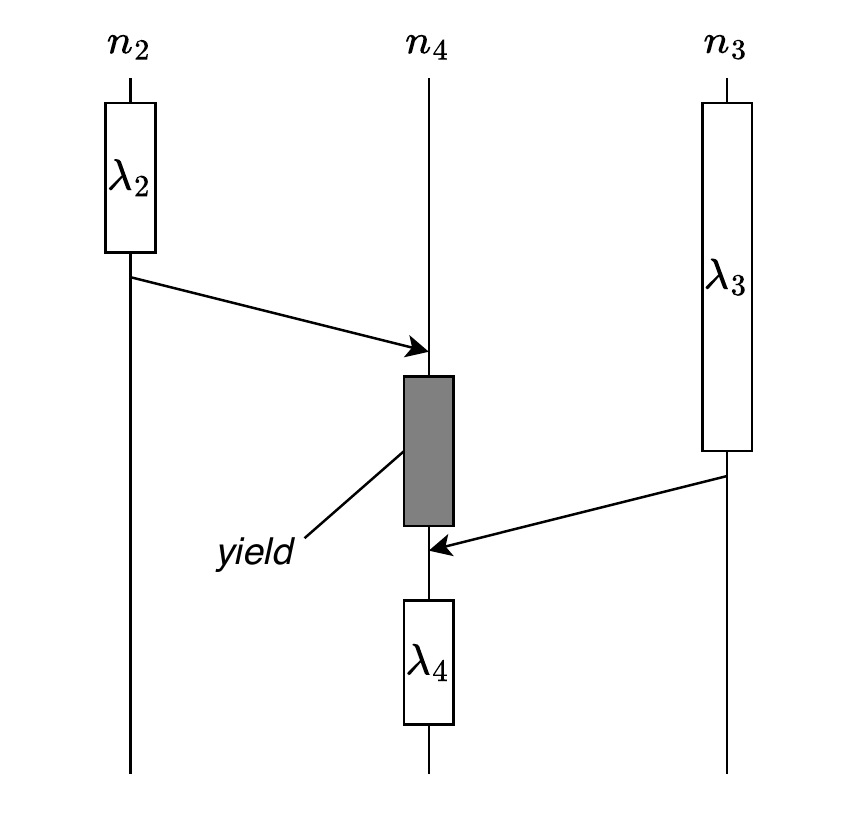}{Example of execution of $\lambda_4$ with yield in the
application in \rfig{dag-1}.}

First, it can happen that a task has more than one
input, e.g., $\lambda_4$ in \rfig{dag-1}: in this case, both $\lambda_2$
and $\lambda_3$ want to execute $\lambda_4$ at the end of their respective
tasks, so the whole notion of
``every function executes the next one'' is not as well-defined as with a
chain of functions.
We address this point by introducing the concept of
\textit{asynchronous calls}: when a function terminates, it always invokes
the next function(s), i.e., its direct descendants according to the \ac{DAG},
but this only triggers the execution of a task if all its inputs are available.
If this is not the case, then the output of the predecessor is stored
temporarily on the edge node and the function \textit{yields} (the term is
borrowed from asynchronous programming models and languages).
A graphical illustration of the yield operation can be found in \rfig{dag-6}.
Supporting this pattern increases the complexity of the
serverless platform on the edge nodes, which have to maintain an ephemeral
state for each incomplete operation.
Such asynchronous calls, by themselves, do not solve the problem:
in the example in \rfig{dag-6} we have assumed that $\lambda_2$ and
$\lambda_3$ both invoke $\lambda_4$ on the same edge node $n_4$,
but in general this is not a piece of information that they have:
the serverless platform treats every function call independently
from others, which can result into the execution of the same function
on different edge nodes.
So, for instance, $\lambda_2$ may invoke $\lambda_4$ on edge node $n_x$
($x \neq 4$), which would result in a deadlock, since both the instances of
$\lambda_4$ on $n_x$ and $n_4$ will yield forever waiting for an input
that will never come.
%

To support StateProp/StateLocal
it is necessary that the mapping between functions and edge
nodes is known to all the workers at least during a single execution of a
\ac{DAG} application.
This way, we can make sure that all the workers will invoke the execution of
descendants on the same edge nodes (i.e., $n_4$ from both $\lambda_2$ and
$\lambda_3$ in the previous example).
The main practical consequences are two:
\begin{myinlinelist}
    \item the information on the mapping between functions and edge nodes
    has to be carried along the execution \ac{DAG}, which slightly increases
    the protocol overhead;
    \item there must be a process that is able to ``resolve'' all the
    functions at the time the \ac{DAG} is invoked (e.g., this can be done
    by the client), which can increase the start-up latency.
\end{myinlinelist}

Frustratingly, all this is not sufficient to support StateProp/StateLocal.
Consider again the trivial example in \rfig{dag-1}: both $\lambda_2$ and
$\lambda_3$ depend on the same state $S$.
Irrespective of the relative order, it will be necessary to transfer the
updated state, modified by the first one to be executed (e.g., $\lambda_2$),
to the other one (e.g., $\lambda_3$).
But there is no invocation path between the two, i.e., $\lambda_3$ is not
a descendant of $\lambda_2$ in the \ac{DAG}, which makes it impossible to
rely on the propagation of the state alone.
Therefore, we propose a second modification: rather than embedding the state
in the function arguments (or their references, for StateLocal), every
function accessing a state will send it directly to the next worker
that will use it, according to the state dependency graph and causal
consistency constraints, both already known.
Thus, the workers of a function must be ready to not only
receive asynchronous calls, and temporarily store their arguments,
but to also store updates states, also arriving asynchronously.
We note that these modifications are not required with PureFaaS,
as illustrated in \rsec{dag:purefaas} above, because the client
provides implicit synchronization with that model.

\myfigfulleps{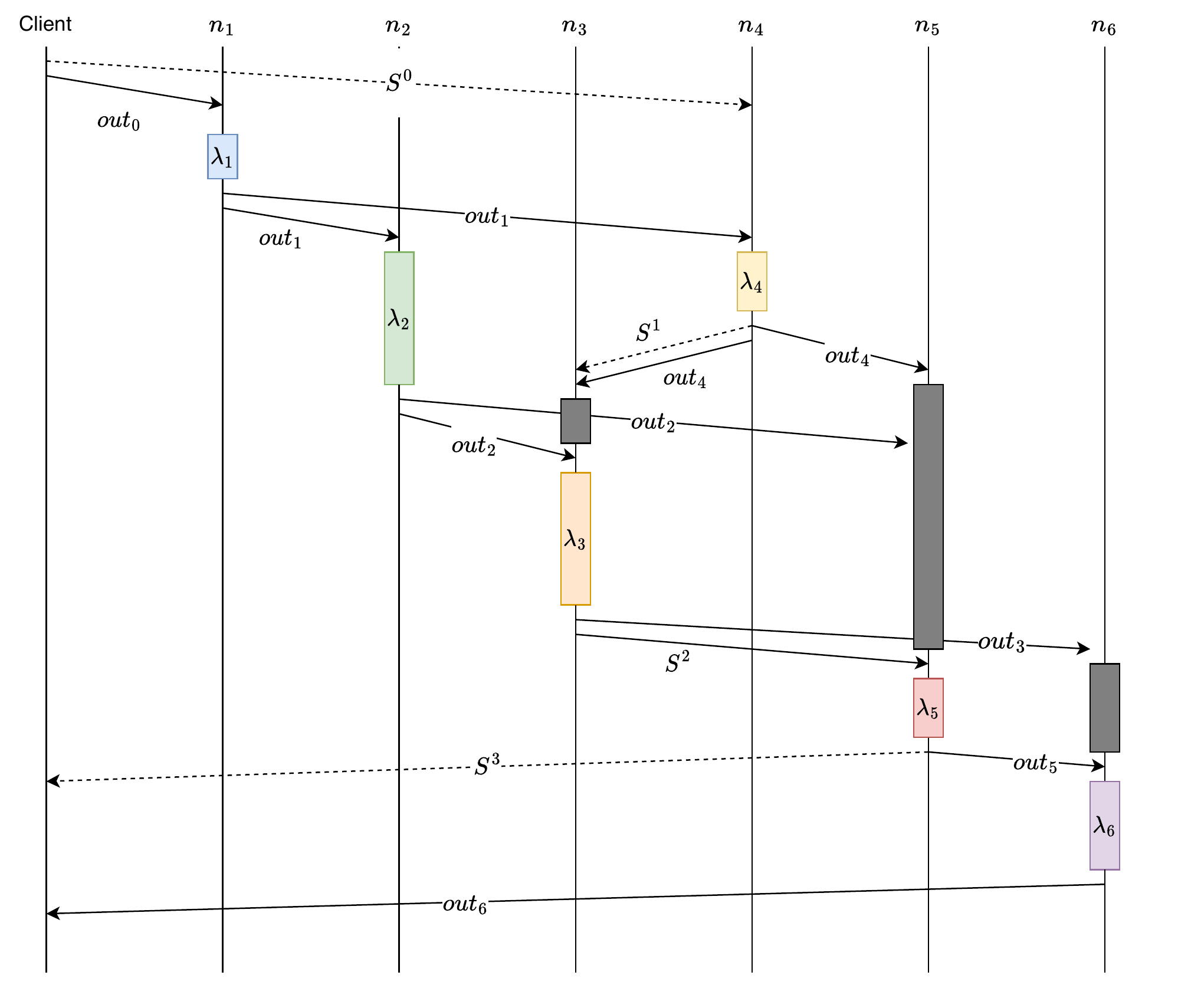}{Sequence diagram of the execution of
the example \ac{DAG} application in \rfig{dag-4} with
StateProp/StateLocal.}

In \rfig{dag-7} we illustrate the sequence diagram with the application
in \rfig{dag-4} of StateProp/StateLocal, the only difference between the
two being that with StateLocal the state has to be retrieved from the
last owner (unless the worker executes on the same edge node).
The amount of data transmitted with StateProp then becomes:
\begin{equation}\label{eq:stateprop2}
    D_{sp}^{DAG} = \sum_{i=0}^N deg^+(\lambda_i) out_i +
        \sum_{j=1}^{M} \left(1 + deg(S_j)\right) S_j,
\end{equation}
where $deg^+(\lambda_i)$ is the out-degree of vertex $\lambda_i$, i.e.,
the number of its direct descendants.
For StateLocal, $D_{sp}^{DAG}$ is an upper bound.

\keypoint{StateProp and StateLocal can support \ac{DAG} applications, but the
following major modifications are needed: workers must support asynchronous
function calls, the binding between functions and edge nodes must
be known to all workers during a single \ac{DAG} execution, and the
states cannot be propagated along with the arguments.
Collectively, these changes increase the complexity of the software to be
run on edge nodes and the protocol overhead, \addedb{as well as exacerbate
possible concerns on disclosing proprietary information (see \rsec{cont:ipr}).}}

\addedb{
\rtab{traffic} summarizes the amount of traffic exchanged with all the
schemes.
}

\begin{table*}[tbp]%
\caption{
    Traffic exchanged for the different policies \addedb{(Chain results are from \cite{Cicconetti2021})}.}%
\centering%
{\small%
\addeda{
\begin{tabularx}{\textwidth}{*{2}{p{0.15\textwidth}}X}
    \hline

    \thead{Chain/DAG} &
    \thead{Policy} &
    \thead{Traffic exchanged} \\

    \hline

    \textit{Both} &
    Pure FaaS &
    $= \sum_{i=0}^N out_i + \sum_{i=1}^{N-1} out_i +
    2 \sum_{j=1}^{M} deg(S_j) S_j$ \\

    Chain &
    StateProp &
    $= \sum_{i=0}^N out_i + (N+1) \sum_{j=1}^{M} S_j$ \\

    Chain &
    StateLocal &
    $\le \sum_{i=0}^N out_i + \sum_{j=1}^{M} deg(S_j) S_j$ \\

    DAG &
    StateProp &
    $= \sum_{i=0}^N deg^+(\lambda_i) out_i +
        \sum_{j=1}^{M} \left(1 + deg(S_j)\right) S_j$ \\

    DAG &
    StateLocal &
    $\le \sum_{i=0}^N deg^+(\lambda_i) out_i +
        \sum_{j=1}^{M} \left(1 + deg(S_j)\right) S_j$ \\
    \hline
\end{tabularx}%
}
}%
\label{tab:traffic}%
\end{table*}%

}

\addeda{%
  \section{Performance evaluation}%
  \label{sec:prototype}%
  In this section we illustrate the prototype we have realized of PureFaaS,
StateProp, and StateLocal
(\rsec{prototype:implementation}) and we report the results obtained
in an emulated network (\rsec{prototype:results}), which complement
the simulations experiments in \cite{Cicconetti2021}.

\myssec{Implementation}{prototype:implementation}

We have implemented the execution models in ServerlessOnEdge, which is a
decentralized framework to dispatch stateless \ac{FaaS} functions at the
edge, developed and maintained within our research group.
The software is open source with a permissive MIT license and publicly available
on GitHub\footnote{\url{https://github.com/ccicconetti/serverlessonedge/},
tag $\ge$ 1.2.1.}~\cite{9193994}.
In ServerlessOnEdge the clients request the invocation of
functions via \textit{e-routers}, which play the role of intermediary with the
serverless platforms by forwarding stateless requests to one of many
destinations available depending on the load and network conditions.
For the purpose of evaluating the performance of protocols and algorithms
in controlled and repeatable conditions, we have also implemented so-called
\textit{e-computers}, which emulate serverless platforms with a given
configuration, in terms of computation speed, memory, number of containers, etc.
ServerlessOnEdge uses Google's
gRPC\footnote{\url{https://grpc.io/}\lastaccessed} for communication
among clients, e-routers, and e-computers.
%

\textbf{PureFaaS} was implemented as follows:
\begin{myinlinelist}
    \item the client embeds the required states within the arguments
    at each function invocation;
    \item the e-computers return the embedded states as part of the function
    return value;
    \item multiple functions are invoked if the precedences are met
    (only in a \ac{DAG}).
\end{myinlinelist}
On the other hand, implementing StateProp and StateLocal required more
structural upgrades.
We start with \textbf{StateProp}, which requires any intermediate
e-computer to invoke the next function(s) in the chain/DAG and pass on all
the application's states.
First, we have implemented asynchronous function calls: they return
immediately an empty acknowledgment, while the real output is
provided to the client as an unsolicited response-only message by
the last e-computer in the chain.
Furthermore, an e-computer in our system does not know the destination of the
next function in the chain: to solve this problem, we have installed on every
e-computer a \textit{companion e-router} that is used to dispatch the
function calls generated by its e-computer as part of the function chain
execution.
To obtain consistent performance of StateProp for both chains and
\acp{DAG}, we have implemented the same state propagation mechanism,
even though this means that not all possible combinations of \ac{DAG} and
state dependencies are supported, see \rsec{dag:prop}
(all those in the experiments are feasible).

\myfigfulleps[width=\textwidth]{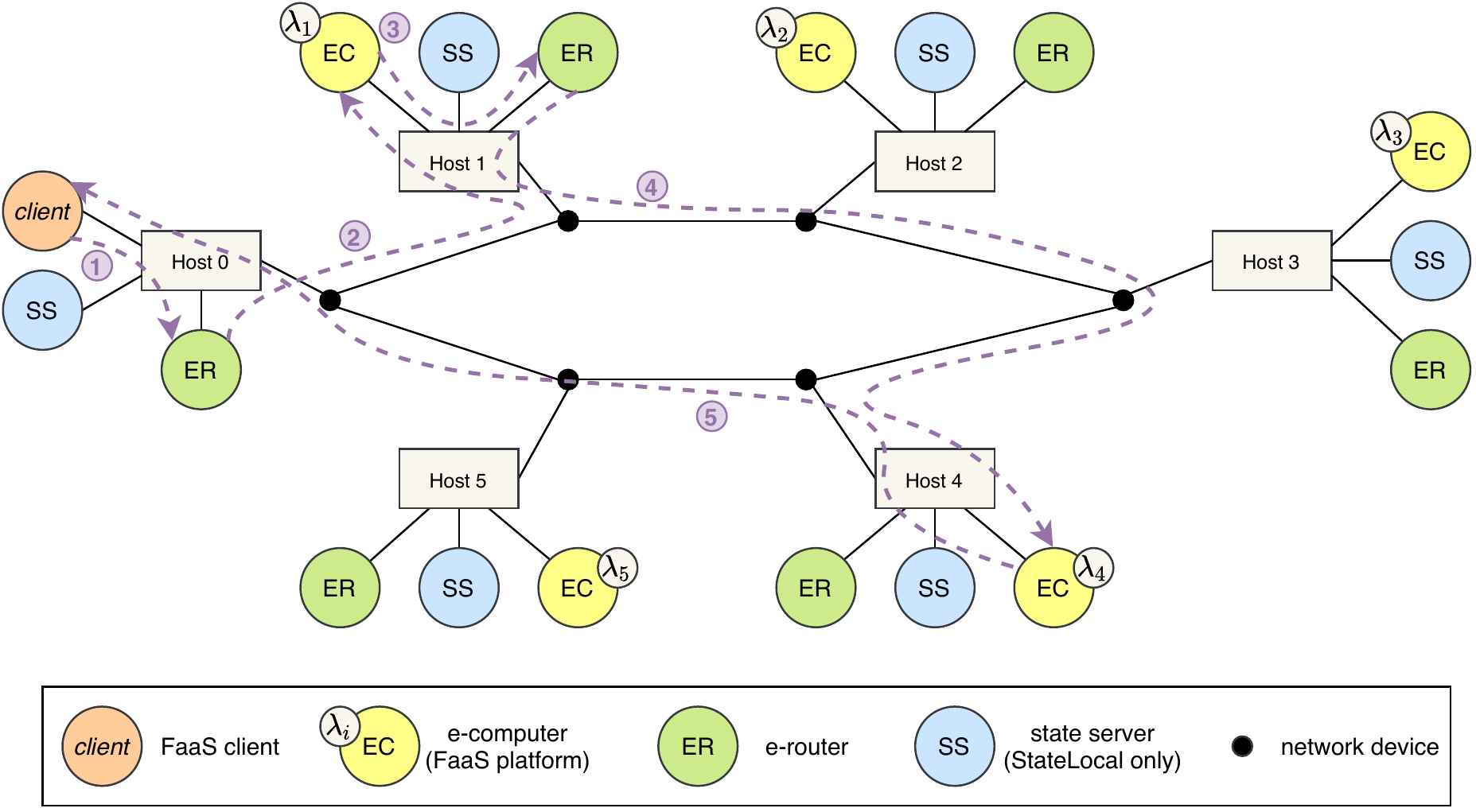}{
    Scenario used for the validation with ServerlessOnEdge.}

An example of invocation of the two-function chain $\{\lambda_1, \lambda_4\}$
is shown in \rfig{soe} in a network with 5 edge nodes (from Host~1 to Host~5),
while the client application is on Host~0.
As can be seen, the e-router on Host~0 is used by the client for the invocation
of the first function in the chain ($\lambda_1$, forwarded to Host~1), while
the e-router on Host~1 receives the next invocation to $\lambda_4$ (forwarded
to Host~4).
The e-computer on Host~4 does not need to go through its companion e-router
as it can send the final response to the client on Host~0.
The system messages had to be modified so that the chain or \ac{DAG}
is embedded in every function request, along with the callback
end-point for the final response.

Finally, \textbf{StateLocal} required the same upgrades as StateProp and
a few others:
\begin{myinlinelist}  
    \item the system messages had to support remote states,
    i.e., states that are not embedded in the function call/response,
    but only referenced indirectly (in our case by their name and
    an end-point);

    \item the states are managed by new components called
    \textit{state servers}, which are simple in-memory \acp{KVS}
    co-located with each e-computer and client, as shown in \rfig{soe};

    \item the flow of messages is exactly the same as with StateProp,
    but at each function invocation the e-computer retrieves the
    remote states needed and then copies them into its local state
    server; to do this, the state dependencies were also embedded
    in the function invocation request messages.
\end{myinlinelist}

\myssec{Results}{prototype:results}

We have used the prototype implementation to carry out a campaign of
experiments in an emulated environment, using
mininet\footnote{\url{http://mininet.org/}\lastaccessed}
to reproduce the topology illustrated in \rfig{soe} above.
The experiments are reproducible by means of the scripts published
in the ServerlessOnEdge GitHub repository (experiment numbers:
\texttt{400} and \texttt{401}), which also includes pointers to
the raw results obtained in the research group servers.
Since we are only interested in measuring the traffic, and its induced
latency, for the different execution models proposed in \rsec{contribution},
we use a single client, i.e., there is no contention on processing
resources.

We have carried out two batches of experiments, respectively with function chains
and \acp{DAG}.
Let us start with \textbf{function chains}:
the client executes back-to-back function chains of constant length $L$ (3 or 6), in
number of functions, where each function is drawn randomly from
${\lambda_1,\ldots,\lambda_5}$, possibly with repetitions.
We assume that the application has $S$ states (3 or 6), where state $s_i$ has size
$(1+i) \times 10~kB$ (0-based indexing); each state depends on a randomly
drawn set of functions, with random cardinality drawn from 0 (no dependencies)
and $L$ (all functions in the chain depend on the state).
The size of the input argument and return value is assumed to be the same
and equal to $A$ (10~kB or 100~kB).
%

    
    
    

\begin{figure*}[tb]
\centering
\includegraphics[width=0.48\textwidth]{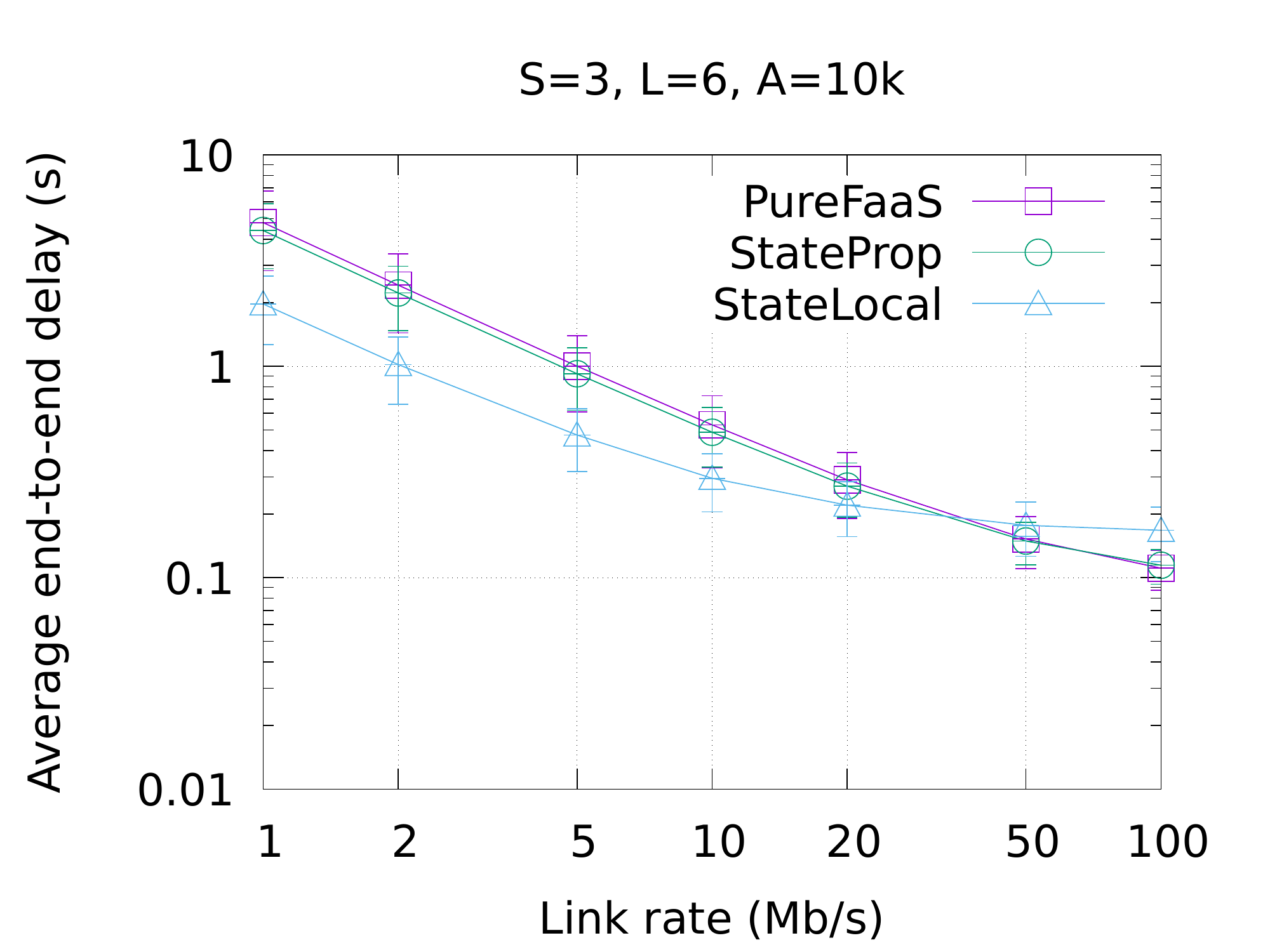}
\includegraphics[width=0.48\textwidth]{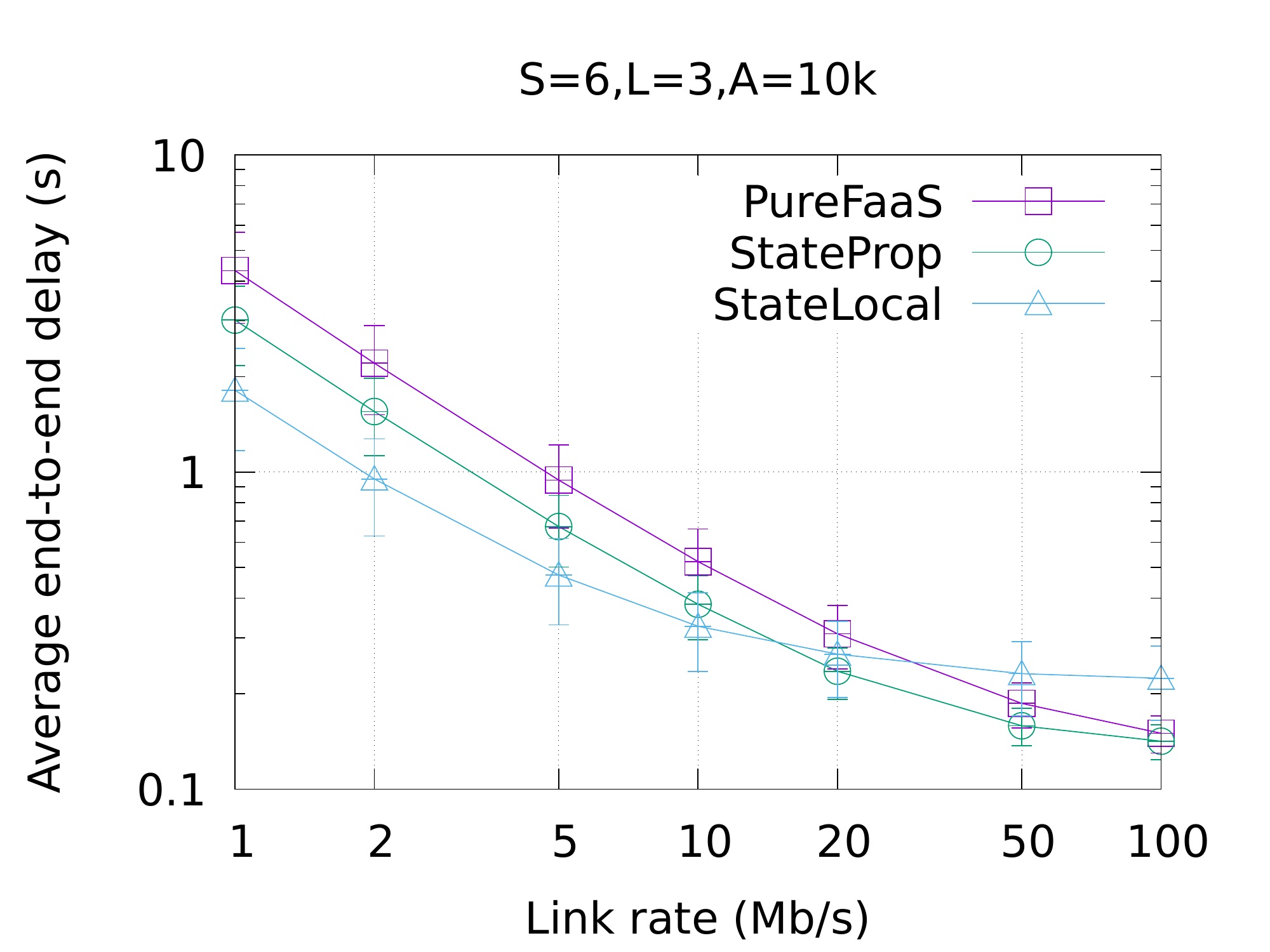}
\includegraphics[width=0.48\textwidth]{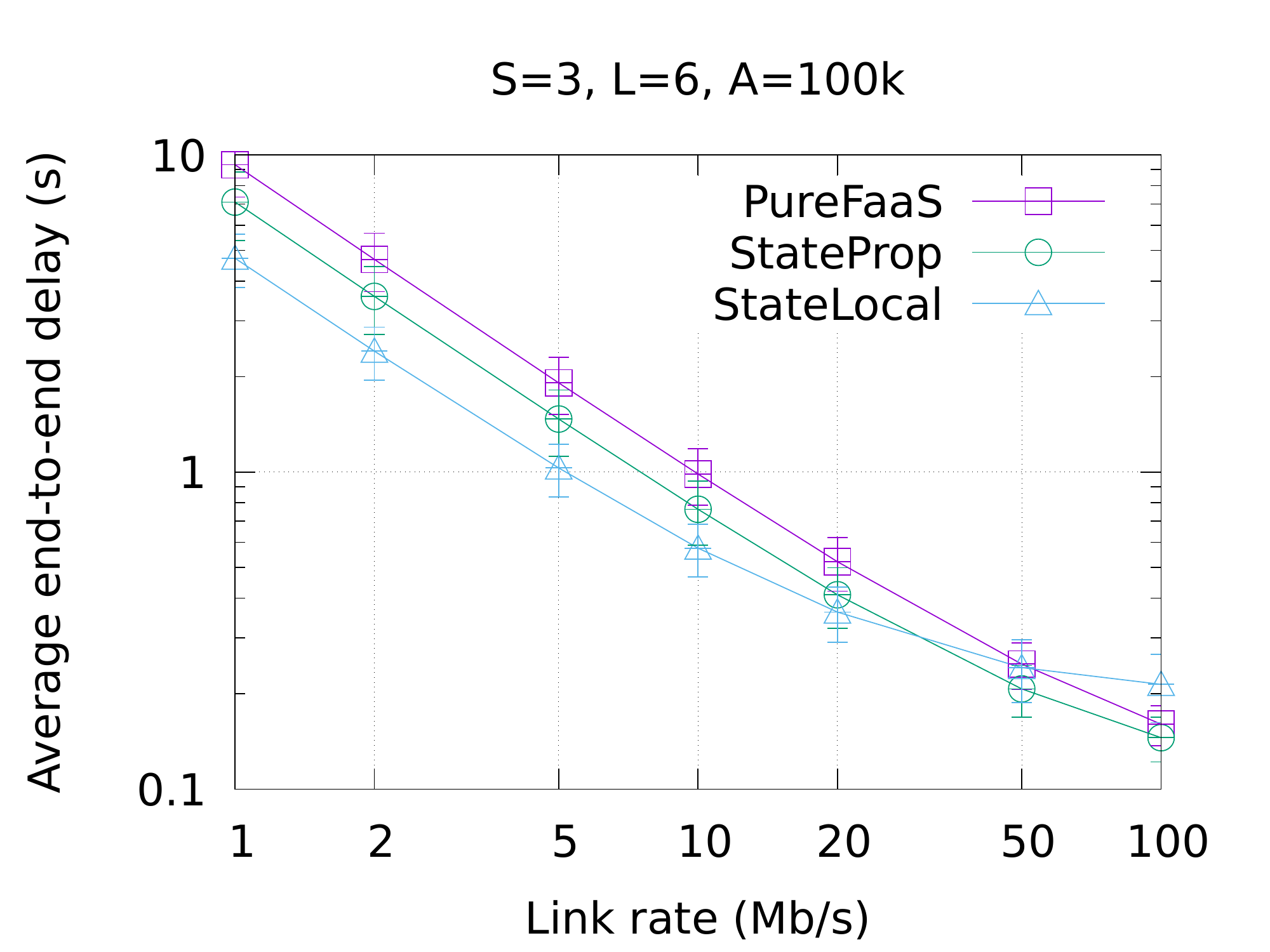}
\includegraphics[width=0.48\textwidth]{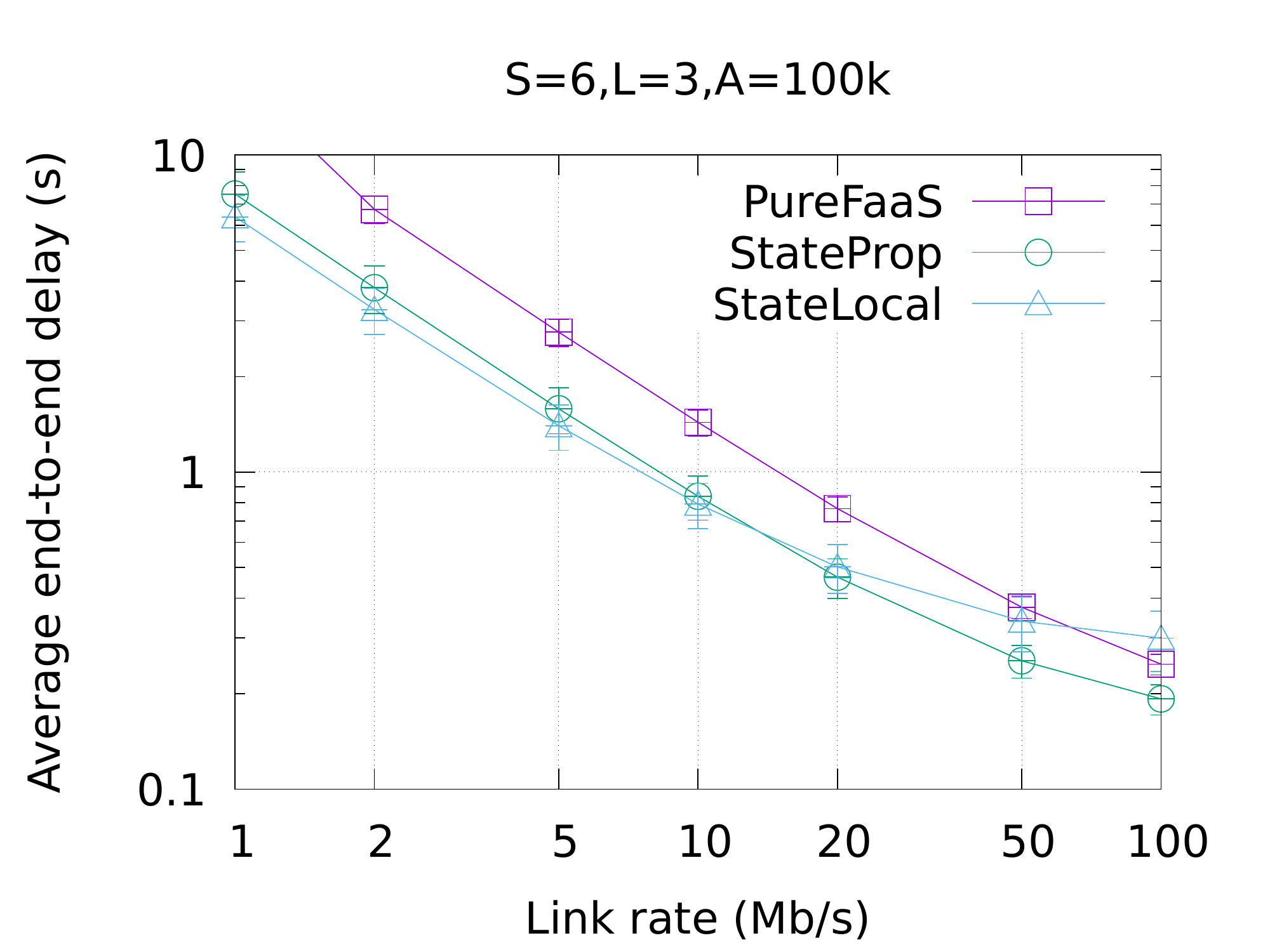}
\vspace{-1em}%
\caption{Function chains: Average end-to-end delay.
}%
\vspace{-1em}%
\label{fig:chain-delay}%
\end{figure*}

In \rfig{chain-delay} we compare the average end-to-end delay obtained with
the execution models in all the scenarios separately, as the link rate between
network devices increases from 1~Mb/s to 100~Mb/s.
Note that the results are plotted in logarithmic scale in both axes.
In the top left plot, PureFaaS and StateProp are almost overlapping: this is
because the size of both states and argument is relatively small.
Instead, until the link rate is below 20~Mb/s, StateLocal has a
much lower average delay, thanks to its wiser only-as-needed transfer of states. 
However, with higher link rates, the advantage diminishes progressively
until the delay becomes higher than that of PureFaaS/StateProp with 50~Mb/s
and 100~Mb/s link rates: at such high connectivity rates, the data transfer
becomes comparable with (or higher than) the time to establish the TCP
connections to retrieve/update the states.
This disadvantage of StateLocal could be reduced by employing persistent
TCP connections towards the state servers or using a connection-less
protocol, such as QUIC\footnote{QUIC: A UDP-Based Multiplexed and
Secure Transport --
\url{https://datatracker.ietf.org/doc/html/rfc9000}\lastaccessed.}.
In the opposite scenario, i.e., bottom right plot in \rfig{chain-delay},
the performance with StateProp and StateLocal are comparable, except for
high link rates: this is because the chains are shorter than in the other
scenario and the data transfer is dominated by the input/output argument,
which is treated the same by the two schemes.
The other cases, i.e., top right and bottom left in \rfig{chain-delay},
are intermediate, with StateLocal achieving a better performance for all
slow link rates, and PureFaaS always lying on top of StateProp.

\keypoint{
    With function chains, when taking into account realistic protocol
    overheads, there is a trade-off between keeping the state local
    to edge nodes (StateLocal) and embedding it into function
    invocations, depending on the state size and network speed.
    StateProp always performs better than PureFaaS.
}

\myfigfulleps[width=\textwidth]{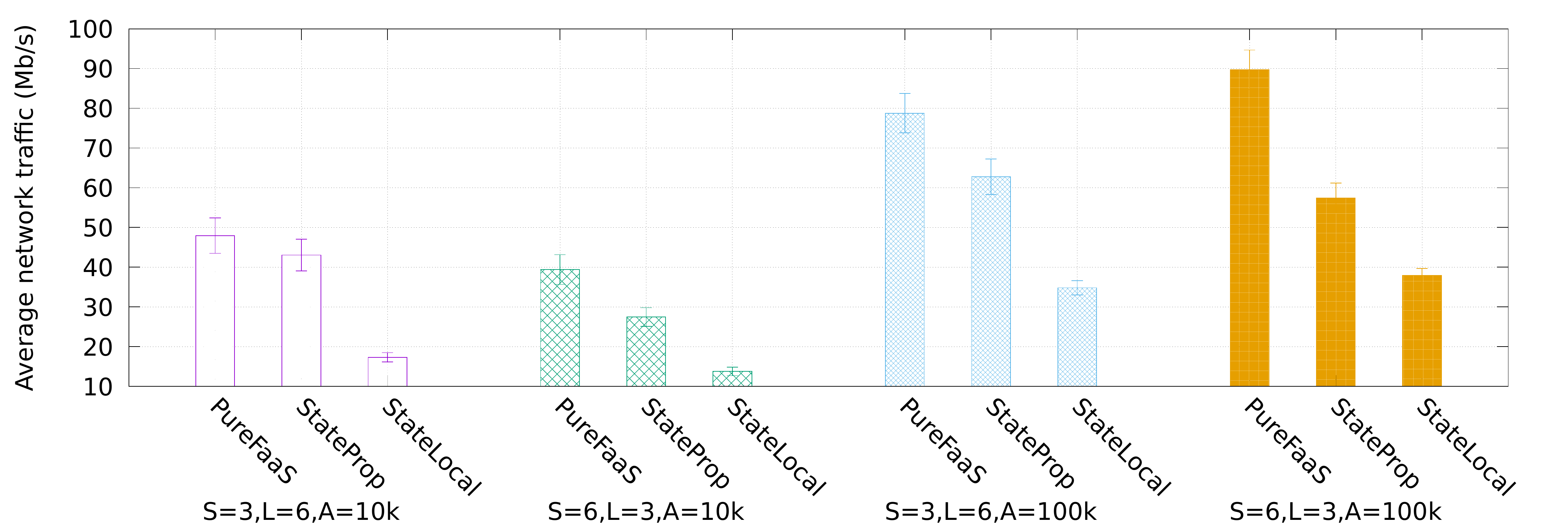}{
    Function chains: Network traffic, with link rate 100~Mb/s.}

We then provide a direct measure of the overhead in
\rfig{chain-tpt-histo-ci} by showing the average network traffic
in all the scenarios, for the link rate 100~Mb/s, which is the one
where StateLocal exhibits worst performance.
As can be seen (in linear scale in this plot) the traffic generated with
PureFaaS is always greater than that generated with StateProp, which in turn
is always greater that that with StateLocal.
We note that, unlike our previous results in~\cite{Cicconetti2021},
the data reported here include all protocol overhead, since the
traffic is measured on the ports of the emulated network switches.
The advantage of StateLocal is more prominent with a smaller argument size,
i.e., with $A = 10~kB$, but it is significant in all cases.

\keypoint{
    With function chains, even with a high network speed, StateLocal
    has a significantly lower overhead than the other schemes, in
    terms of the traffic rate required, but this not always
    translates into a lower end-to-end latency.
}

\begin{figure*}[tb]
    \centering
    \includegraphics[width=0.32\textwidth]{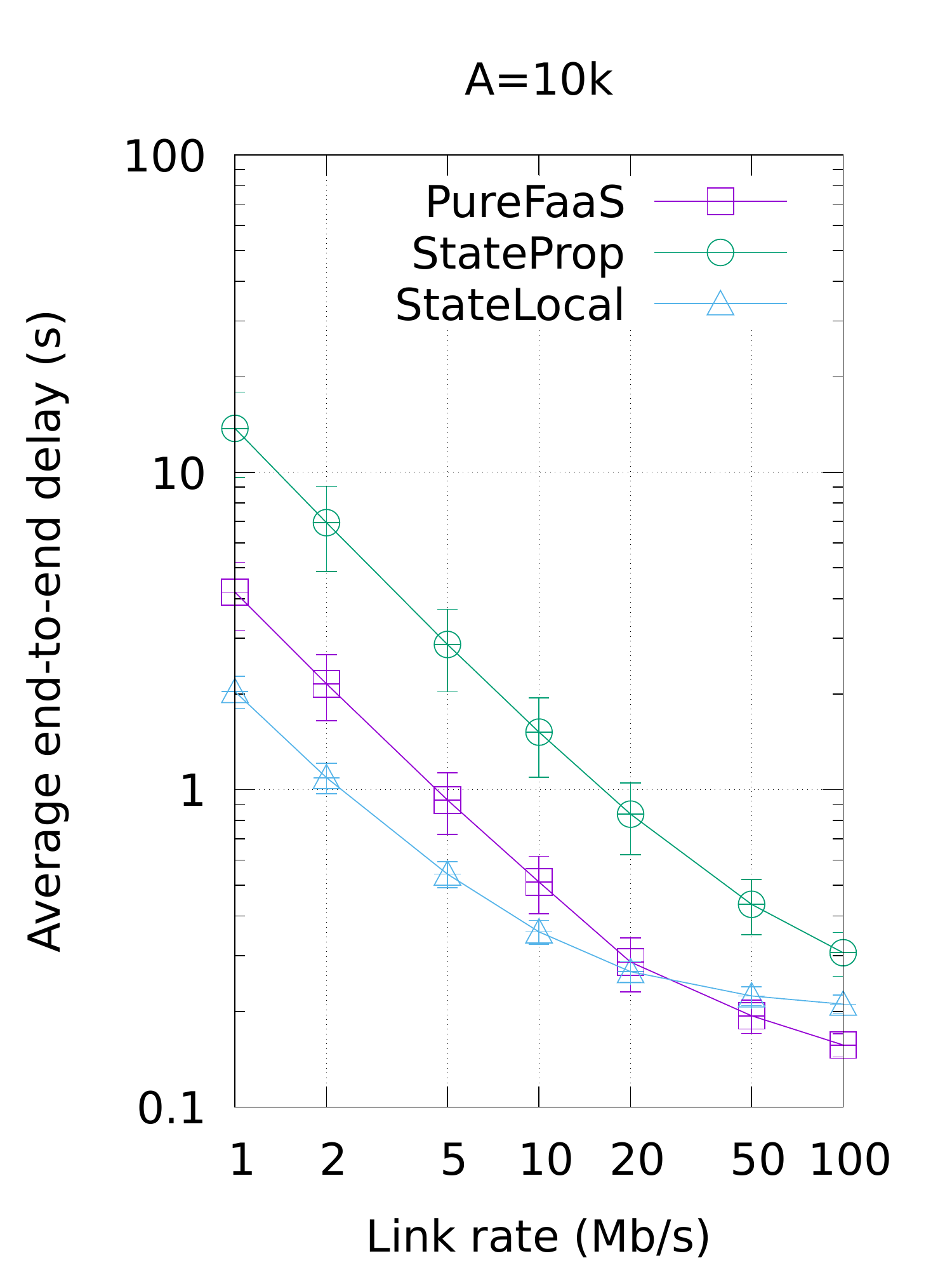}
    \includegraphics[width=0.32\textwidth]{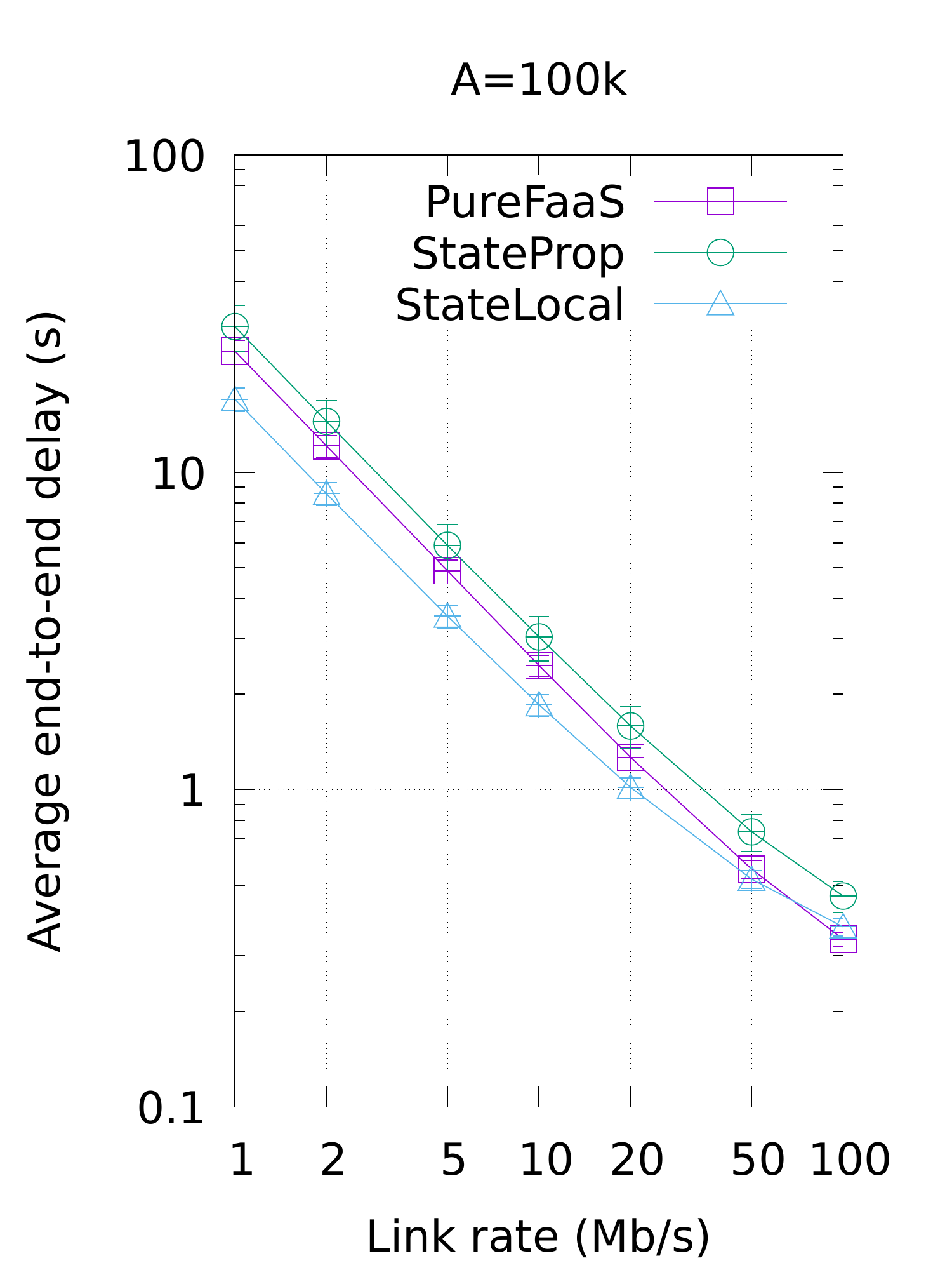}
    \includegraphics[width=0.32\textwidth]{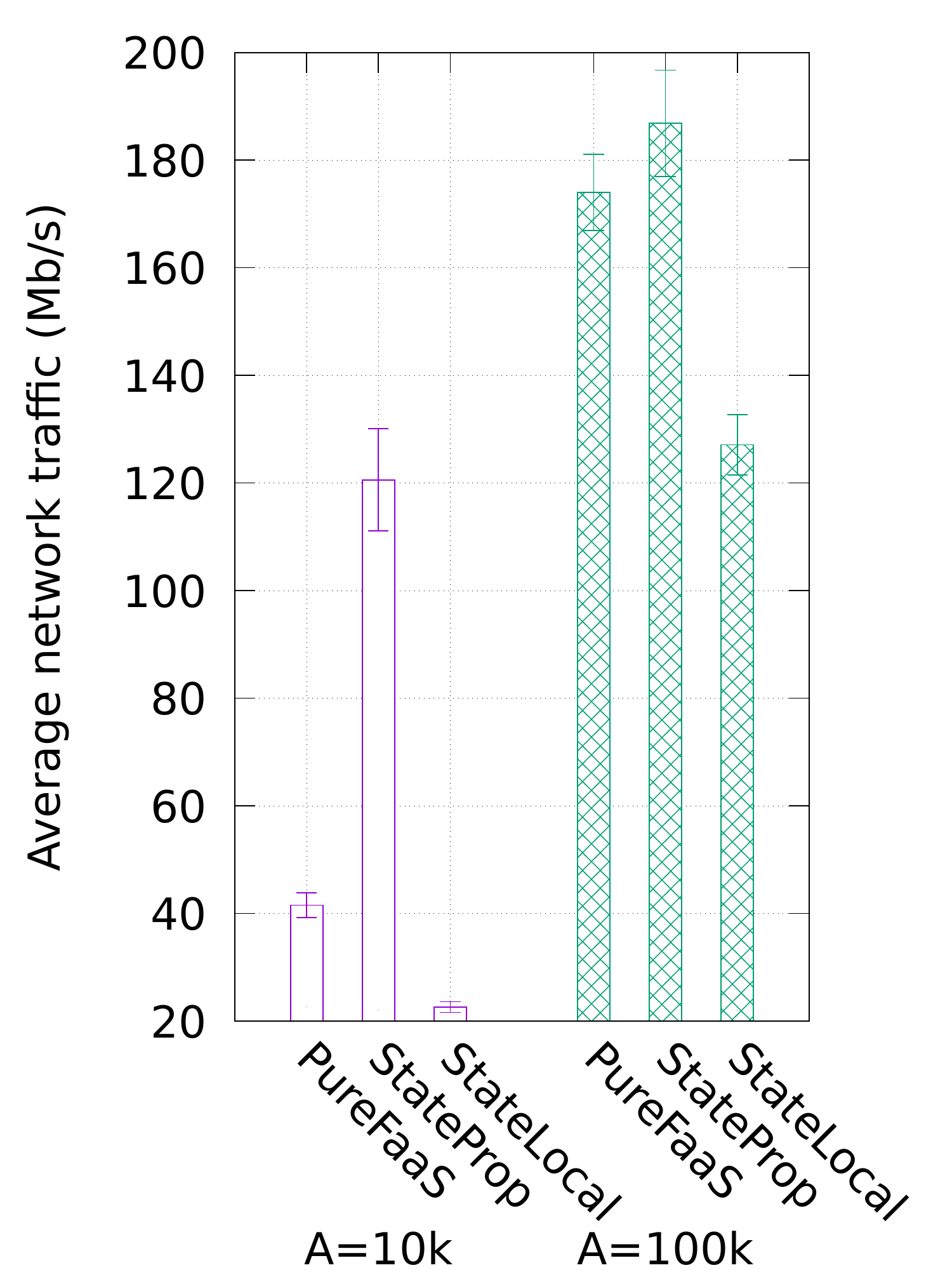}
    \vspace{-1em}
    \caption{\acp{DAG}: Average delay, with variable link rates, and network traffic, with link rate 100~Mb/s.}
    \vspace{-1em}
    \label{fig:dag-proto-results}
\end{figure*}

We now move to the \textbf{\ac{DAG}} case, for which we
considered applications made of a sequence of stages, each with a
\textit{branch} function that spawns multiple stateless calls
followed by a stateful \textit{collect} task.
\addedb{
This structure is very typical of \ac{ML} applications, which are
today dominant in cloud and edge environments: this was confirmed
by the study \cite{Tian2019}, where the authors have synthesized an artificial
generator of workloads that captures accurately, in a statistical
sense, the behavior of real-life applications in the wild.
}
In the results below, we have used 3 stages with 5 branches per stage.
The functions are selected randomly among those on the edge nodes,
and the state dependencies are also random using the same approach as with
chains.
In \rfig{dag-proto-results} we show the average delay, with increasing link rate
from 1~Mb/s to 100~Mb/s (with a log-log plot), as well as the network traffic
only for the link rate 100~Mb/s.
As above, we have used two argument sizes: $A=10~kB$ and
$A=100~kB$.
Unlike with function chains, in this scenario we find that PureFaaS outperforms
StateProp in all conditions, with the advantage being more prominent with
a smaller argument size.
This is because the total number of functions called is much higher than
in the chain scenario, which penalizes significantly the embedding of all the
states in invocations and responses.
Such a fee is not paid by StateLocal, which only transfers \textit{references} to
states and performs best in all conditions except with very high link rates,
due to the overhead of state retrieve/update operations, as already discussed.

\keypoint{
    With a high number of functions in \acp{DAG}, state
    propagation is only effective if references are carried within the
    function invocations and responses.
}

}

  \section{Conclusions}%
  \label{sec:conclusions}%
  In this paper we have explored the support of
stateful applications on serverless platforms distributed on
edge nodes.
We have focused on the problem of transferring the state along an
invocation of functions in chain \addeda{and \ac{DAG} workflows},
and we have identified three alternative schemes, with different
characteristics.
%
%
\addeda{We have developed a prototype implementation to prove
the feasibility of our approaches and to measure performance with
realistic protocol overheads.}
The results have shown that propagating the state along the chain
of function invocations can reduce significantly the communication
overhead.
\addeda{
This leads to lower end-to-end application latency, especially
with limited connectivity.
However, with large \ac{DAG} workflows, embedding the state for
propagation is not effective anymore: in these cases it becomes mandatory
to store the states locally on edge nodes and carry their references instead.
}
%
%


\section*{Acknowledgment}

This work was partially supported by the European Union's Horizon
2020 research and innovation programme under grant agreement No
957337, project MARVEL.

\begin{acronym}
  \acro{3GPP}{Third Generation Partnership Project}
  \acro{5G-PPP}{5G Public Private Partnership}
  \acro{AA}{Authentication and Authorization}
  \acro{ADF}{Azure Durable Function}
  \acro{AI}{Artificial Intelligence}
  \acro{API}{Application Programming Interface}
  \acro{AP}{Access Point}
  \acro{AR}{Augmented Reality}
  \acro{BGP}{Border Gateway Protocol}
  \acro{BSP}{Bulk Synchronous Parallel}
  \acro{BS}{Base Station}
  \acro{CDF}{Cumulative Distribution Function}
  \acro{CFS}{Customer Facing Service}
  \acro{CPU}{Central Processing Unit}
  \acro{DAG}{Directed Acyclic Graph}
  \acro{DHT}{Distributed Hash Table}
  \acro{DNS}{Domain Name System}
  \acro{ETSI}{European Telecommunications Standards Institute}
  \acro{FCFS}{First Come First Serve}
  \acro{FSM}{Finite State Machine}
  \acro{FaaS}{Function as a Service}
  \acro{GPU}{Graphics Processing Unit}
  \acro{HTML}{HyperText Markup Language}
  \acro{HTTP}{Hyper-Text Transfer Protocol}
  \acro{ICN}{Information-Centric Networking}
  \acro{IETF}{Internet Engineering Task Force}
  \acro{IIoT}{Industrial Internet of Things}
  \acro{IPP}{Interrupted Poisson Process}
  \acro{IP}{Internet Protocol}
  \acro{ISG}{Industry Specification Group}
  \acro{ITS}{Intelligent Transportation System}
  \acro{ITU}{International Telecommunication Union}
  \acro{IT}{Information Technology}
  \acro{IaaS}{Infrastructure as a Service}
  \acro{IoT}{Internet of Things}
  \acro{JSON}{JavaScript Object Notation}
  \acro{K8s}{Kubernetes}
  \acro{KVS}{Key-Value Store}
  \acro{LCM}{Life Cycle Management}
  \acro{LL}{Link Layer}
  \acro{LTE}{Long Term Evolution}
  \acro{MAC}{Medium Access Layer}
  \acro{MBWA}{Mobile Broadband Wireless Access}
  \acro{MCC}{Mobile Cloud Computing}
  \acro{MEC}{Multi-access Edge Computing}
  \acro{MEH}{Mobile Edge Host}
  \acro{MEPM}{Mobile Edge Platform Manager}
  \acro{MEP}{Mobile Edge Platform}
  \acro{ME}{Mobile Edge}
  \acro{ML}{Machine Learning}
  \acro{MNO}{Mobile Network Operator}
  \acro{NAT}{Network Address Translation}
  \acro{NFV}{Network Function Virtualization}
  \acro{NFaaS}{Named Function as a Service}
  \acro{OSPF}{Open Shortest Path First}
  \acro{OSS}{Operations Support System}
  \acro{OS}{Operating System}
  \acro{OWC}{OpenWhisk Controller}
  \acro{PMF}{Probability Mass Function}
  \acro{PU}{Processing Unit}
  \acro{PaaS}{Platform as a Service}
  \acro{PoA}{Point of Attachment}
  \acro{QoE}{Quality of Experience}
  \acro{QoS}{Quality of Service}
  \acro{RPC}{Remote Procedure Call}
  \acro{RR}{Round Robin}
  \acro{RSU}{Road Side Unit}
  \acro{SBC}{Single-Board Computer}
  \acro{SDN}{Software Defined Networking}
  \acro{SJF}{Shortest Job First}
  \acro{SLA}{Service Level Agreement}
  \acro{SMP}{Symmetric Multiprocessing}
  \acro{SoC}{System on Chip}
  \acro{SRPT}{Shortest Remaining Processing Time}
  \acro{SPT}{Shortest Processing Time}
  \acro{STL}{Standard Template Library}
  \acro{SaaS}{Software as a Service}
  \acro{TCP}{Transmission Control Protocol}
  \acro{TSN}{Time-Sensitive Networking}
  \acro{UDP}{User Datagram Protocol}
  \acro{UE}{User Equipment}
  \acro{URI}{Uniform Resource Identifier}
  \acro{URL}{Uniform Resource Locator}
  \acro{UT}{User Terminal}
  \acro{VANET}{Vehicular Ad-hoc Network}
  \acro{VIM}{Virtual Infrastructure Manager}
  \acro{VR}{Virtual Reality}
  \acro{VM}{Virtual Machine}
  \acro{VNF}{Virtual Network Function}
  \acro{WLAN}{Wireless Local Area Network}
  \acro{WMN}{Wireless Mesh Network}
  \acro{WRR}{Weighted Round Robin}
  \acro{YAML}{YAML Ain't Markup Language}
\end{acronym}

\end{document}